\documentclass[5p,twocolumn,10pt,times]{elsarticle}

\usepackage{lineno,hyperref}
\journal{Computer Aided Design}

\newcommand{\etal}{et al.~}
\newcommand{\ie}{\textit{i}.\textit{e}.,~}
\newcommand{\eg}{\textit{e}.\textit{g}.,~}

\usepackage{multirow}
\usepackage[ruled]{algorithm2e} 
\usepackage{amsmath}
\usepackage{amsfonts}
\usepackage{xcolor}
\usepackage{caption}
\usepackage[draft]{pgf}
\usepackage{caption}
\usepackage{subcaption}
\usepackage{esint}
\usepackage{amssymb}
\usepackage{wrapfig} 
\usepackage{algpseudocode}

\begin{document}
\baselineskip11pt

\begin{frontmatter}

\title{Sliding Basis Optimization for Heterogeneous Material Design}

 \author{Nurcan Gecer Ulu, Svyatoslav Korneev, Erva Ulu, Saigopal Nelaturi}
\address{Palo Alto Research Center (PARC), 3333 Coyote Hill Rd, Palo Alto, California}

\begin{abstract}

We present the \emph{sliding basis} computational framework to automatically 
synthesize heterogeneous (graded or discrete) material fields for parts 
designed using constrained optimization. Our framework uses the fact that any spatially varying material field over a given domain may be parameterized as a  
weighted sum of the Laplacian eigenfunctions. We bound the parameterization of all material fields using a small set of weights to truncate the Laplacian eigenfunction expansion, which  enables efficient  design space exploration with the weights as a small set of design variables. We further  improve computational efficiency by  using the property that the Laplacian eigenfunctions  form a spectrum and may be ordered from lower to higher frequencies.  Starting the optimization with a small set 
of weighted lower frequency basis functions we iteratively include higher 
frequency bases by sliding a window over the space of ordered basis functions 
as the optimization progresses. This approach allows greater localized control 
of the material distribution as the sliding window moves through higher 
frequencies.  The approach also reduces the number of optimization variables 
per iteration, thus the design optimization process speeds up independent of the 
domain resolution without sacrificing  analysis quality. While our method is 
useful for problems where analytical gradients are available, it is most 
beneficial when the gradients may not be computed easily (\ie optimization 
problems coupled with external black-box analysis) thereby enabling optimization 
of otherwise intractable design problems. The sliding basis framework is 
independent of any particular physics analysis, objective and constraints, 
providing a versatile and powerful design optimization tool for various 
applications. We demonstrate our approach on graded solid rocket fuel design and 
multi-material topology optimization applications and evaluate its performance.

\end{abstract}


\end{frontmatter}


\section{Introduction}

Multi-material (heterogeneous) structures show great potential for superior 
product performance compared to homogeneous material designs. The advantages of 
 heterogeneous material structures such as fiber reinforced, metal matrix, and 
ceramic matrix composites are already clear with several applications in 
aerospace engineering, construction, transportation, medical,  and defense 
industries. In some applications (such as spacecraft engineering) where 
traditional composite materials may fail prematurely via delamination and other 
mechanisms, functionally graded materials characterized by gradual transitions 
in material compositions and microstructure can improve performance by 
improving mechanical properties and avoiding hard interfaces between materials.

Realizing this potential, a variety of multi-material AM technologies have 
been developed for different material types including 
polymers~\cite{StratasysObjet}, metals~\cite{Insstek} and 
ceramics~\cite{LiLiquin:2017}. These multi-material AM technologies are 
becoming commercially available and are being used in real world applications 
~\cite{Bandyopad:2018}. However, design tools that can take full 
advantage of such manufacturing capabilities are missing. Manually designing 
heterogeneous material structures is a difficult and tedious task even for 
experienced engineers and designers because multi-material AM technologies can 
enable voxel-level control and therefore create a vast design space. For example 
consider a heterogeneous material design problem with $m$ discrete materials in 
a discretized domain with $n_e$ elements; the resulting design space has $n_e^{m}$ possible combinations 
of materials that may be distributed. The combinatorics are intractable very 
quickly even for low-resolution three dimensional heterogeneous discrete 
materials with voxel level control. When reasoning about weighted combinations 
of materials there are infinitely many combinations per voxel. The advantages 
of heterogeneous graded materials in aerospace and defense applications with 
complex interacting physics further implies the design space exploration must be 
efficiently coupled with domain-specific solvers to synthesize novel 
heterogeneous material designs.

\begin{figure*}[]
  \centering  
  \includegraphics[trim = 0in 0in 0in 0in, clip, width = 
\textwidth]{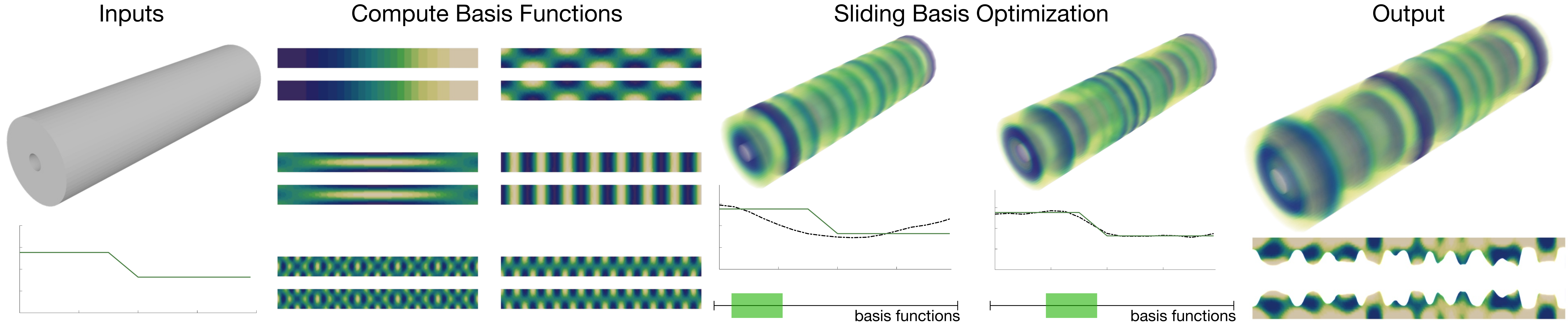}
  \caption{Overview of our approach. Given an input domain and target goals 
(e.g. thrust profile when designing a graded material solid rocket fuel), our 
algorithm optimizes the material distribution in the domain to find a 
configuration that matches the target goals. In the preprocessing step, we 
compute the Laplacian basis functions on the input domain. Then, our sliding 
basis optimizer adjusts the weights of a small number of active basis functions 
at a time. As the optimization progresses, we slide the active set towards the 
higher frequencies until convergence is achieved.}
  \label{fig:overview}
\end{figure*}

 In this paper, we present a novel method for optimizing heterogeneous material 
distributions given a prescribed set of design 
goals~(Figure~\ref{fig:overview}). In particular, we address inverse problems 
where the objective and constraints are coupled with a `black-box' physical 
analysis whose implementation details are unknown.  Our approach may be 
contrasted with the prominent automated synthesis technique of topology 
optimization~\cite{Bendsoe:2004}. Typically, topology optimization approaches 
rely on the idea that gradients related to simulation variables can be computed 
analytically. While these analytical gradients and corresponding adjoint 
variables are well defined for a certain set of problems such as simple linear 
elasticity problems, deriving analytical gradients may be problematic 
or costly for applications involving complex multi-physics especially when 
depending on external solvers for the analysis. Often, such external analysis 
tools do not provide the analytical gradient components that are essential in 
the traditional topology optimization processes.

When analytical gradients are not available, optimization
approaches either use numerical gradients or employ 
stochastic sampling methods such as genetic algorithms~\cite{Davis:1991} or simulated 
annealing~\cite{Kirkpatrick:1983}. These approaches do not scale well 
with increasing number of optimization variables \ie the size of the material 
distribution field, because the introduction of new optimization variables 
requires additional analysis calls. We address this challenge by using a 
reduced order approach that controls the material distribution in a 
high-resolution analysis domain with a small number of design parameters. The 
key underlying idea is that any heterogeneous material field 
$\boldsymbol{\mathcal{F}}$ viewed as a function over a compact domain $\Omega$ 
can be parameterized as a weighted combination of the eigenfunctions of the 
Laplace operator $\Delta = \nabla \cdot \nabla$ defined over $\Omega$ (with 
Dirichlet boundary conditions). Every Laplacian eigenfunction $f$ 
satisfies the 
relation $\Delta f = -\lambda f$, and the set of all such $\Omega$-specific 
Laplacian 
eigenfunctions form a complete orthonormal basis $\{ \boldsymbol{e_i} \}$ for 
the function space 
$L^2(\Omega)$. Therefore we may write
\begin{equation}
\boldsymbol{\mathcal{F}} = \sum_i \boldsymbol{w_i} 
\boldsymbol{e_i}.
\end{equation}

In our approach, the design variables are the weights $\boldsymbol{w_i}$ 
applied to the pre-computed $\boldsymbol{e_i}$ for a given $\Omega$. In a 
discrete representation of $\Omega$ (e.g. as a mesh) there are finitely many 
$\boldsymbol{e_i}$ determined by the number of mesh elements. Using the 
eigenfunction expansion and treating  weights $\boldsymbol{w_i}$ as design 
variables, we see there are as many design variables as the size of the 
material field (mesh elements), so this expansion simply amounts to a basis 
change. However, the power of the eigenfunction expansion emerges when we 
truncate the basis to create  a much smaller set of optimization variables 
compared to the size of the field in the discretized $\Omega$. The 
$\boldsymbol{e_i}$ have the \emph{spectral property} 
\cite{sorkine2005laplacian}, i.e. they form a spectrum and can be ordered 
from low to high frequencies, and the higher frequency eigenfunctions have support over very small features. Truncating
the number of basis elements removes high 
frequency eigenfunctions from the material field parameterization, so the 
computational advantage of reducing the design variables (weights) at the 
expense of local material field variation must be traded off carefully. Our goal 
is to arrive at a truncation that is sufficient to parameterize and represent 
the material field that optimizes the given objective and constraints. 

To avoid the trial and 
error in selecting a smaller set of basis functions to efficiently explore the design space, we introduce the \emph{sliding basis 
optimization}. The key observation we make in this work is that the spectral 
properties of the Laplacian eigenfunction basis allows us to capture material
field variation over increasingly local features by sliding (i.e. incrementally 
moving) towards the 
higher frequency basis functions as the optimization progresses. Our method can 
be used with both numerical gradients and stochastic optimization approaches 
incorporating commodity optimizers. The method provides a flexible and powerful 
mechanism for material distribution design that can be easily applied to a 
variety of problems. We demonstrate two example applications. First, we apply it 
to graded solid rocket fuel design such that the optimized fuel distribution 
results in the target thrust profile when it burns. Second, we show its 
performance on multi material topology optimization problem where the objective 
is to minimize the compliance of the resulting design. For both applications, 
the analysis component is treated as black-box to demonstrate the effectiveness 
of the approach.

The main contributions of the presented work are:
\begin{itemize}
\item an optimization technique we call sliding basis optimization to explore 
parameterized design space efficiently and utilize minimal set of basis to 
achieve design requirements,
\item application of the spectral Laplacian basis to practical material design 
problems with prescribed material bounds,
\item enabling optimization of material distributions for new applications 
coupled with black-box analysis .
\end{itemize}

\section{Related Work}

\subsection{Heterogeneous Material Design}
Many topology optimization algorithms 
have been developed to handle (discrete and graded) heterogeneous material 
distributions. Among all of them, the most popular approach is solid isotropic 
material with penalization~(SIMP) method due to its conceptual simplicity and 
practicality~\cite{Bendsoe:2004}. Recently, an ordered multi-material SIMP 
approach has been presented in~\cite{Zuo:2017multi} that eliminates the 
dependence of computational cost to the number of materials considered. SIMP 
approaches have also been extended to structural optimization of laminate 
composites~\cite{Blasques:2014composite}. The level set approach to topology 
optimization has also been used to design shapes with discrete
heterogeneous materials~\cite{Wang:2004levelset, mirzendehdel2015pareto}. 
Recently, the SIMP and level set topology optimization approaches have been 
extended to graded material design problems. Example applications include 
compliant mechanism design~\cite{Conlan:2019stress}, auxetic material 
design~\cite{Vogiatzis:2017multi}, thermal 
applications~\cite{Vaissier:2019graded} and cellular structure 
design~\cite{Li:2018graded}. In this paper, we build on these heterogeneous 
material topology optimization approaches. Our reduced order method can be 
easily applied to various problems involving different analysis, objectives and 
constraints including but not limited to the ones mentioned above. Our approach 
is complementary to existing gradient based topology optimization methods. 
The parameterization we use has a linear relationship between the weights of 
the basis function~(\ie optimization variables) and the represented material 
field. This differentiable relationship provides a convenient way to incorporate 
our model reduction approach into existing gradient based methods through a 
simple chain rule multiplication. 

\subsection{Reduced Order Design Optimization}
The popularity of topology optimization as an effective approach to generative design has led to 
a growing interest in order reduction techniques that provide a compelling way 
for reducing the computational complexity in optimization 
problems~\cite{Choi:2019accelerating,Amsallem:2015}. For example, an on-the-fly reduced order model construction method has been presented in~\cite{Gogu:2015improving} for large scale structural topology optimization. Yoon~\etal \cite{Yoon:2010reduction} proposes a model reduction approach to reduce the size of the dynamic stiffness matrix for topology optimization of frequency response problems. While these approaches reduce the complexity of physics analysis, other works focus on reducing the number of design variables in the optimization while preserving the desired simulation accuracy. Guest \etal~\cite{Guest:2010reducing} presents a dimension reduction method for structural topology optimization using Heaviside projection. It defines control points that influence the mesh elements within a predefined radius providing local support over the domain. Transforming design variables of topology optimization into wavelet basis have been explored in~\cite{poulsen2002topology}. Zhou \etal~\cite{zhou2018highly} presents a reduced order topology optimization approach using discrete cosine transform and demonstrate its efficiency on 2D problems. Similarly, a topology optimization method has been developed using fourier representations in the form of discrete cosine transforms for compliance minimization problems in ~\cite{white2018toplogical}. In this work, we present a similar approach in the sense that we transform design variables into a different basis and reduce number of design variables. However, we use Laplacian basis and exploit its spectral properties for efficient space exploration through our sliding basis optimization approach.

Laplacian energy based deformation handles are 
used to manipulate designs using small number of variables for shape 
optimization in~\cite{Ulu:2018coupling}. A Laplacian based order reduction has 
been presented in~\cite{Ulu:2017lightweight} for topology optimization of 
problems with load uncertainties. Driven by similar motivations, our approach 
reduces the dimensionality of the optimization variables rather than the 
analysis solution. We use the Laplacian eigenfunction basis to represent the 
material distribution with small number of variables. Compared to previous 
work, our approach 
addresses a more general class of problems involving graded and multi-material 
design. In addition, our sliding basis optimization allows us to explore a 
larger design space effectively by exploiting the spectral properties of the 
Laplacian eigenfunction basis elements.

\subsection{Laplacian Eigenfunction Basis}
Generalizing the Laplacian to Riemannian manifolds using the tools of discrete exterior calculus leads to the well known Laplace-Beltrami operator. It is known that the eigenfunctions of the Laplacian/ Laplace-Beltrami operator define a Fourier-like basis to perform spectral analysis on manifolds \cite{levy2006laplace}; for example the eigenfunctions of the Laplacian on a sphere yield the spherical harmonics. Computing Laplacian eigenfunctions over a mesh has several applications in geometry processing e.g. u-v parameterization~\cite{Levy:2002least}, shape editing by designing filters in the manifold `frequency domain'  ~\cite{Vallet:2008spectral},
segmentation~\cite{Liu:2004segmentation}, computing deformation fields for mesh 
editing~\cite{sorkine:2004laplacian}, and interactive design for haptics and 
animation~\cite{Xu:2015interactive}, among others. Depending on the structure of the mesh and the specific problem, variants of the discretized Laplacian~\cite{Zhang:2004discrete} are used. For example the area weighted or cotangent weighted Laplacian formulation is beneficial in mesh processing applications when the domain is discretized by a non-uniform triangulation. The spectral mesh processing course~\cite{Levy:2010course}  provides a detailed review of the 
applications enabled by this representation. 

Among the varying definitions of the discretized Laplacian, we  note that considering the volumetric mesh as a graph leads to the definition of the graph Laplacian \cite{sorkine:2004laplacian}. Although different geometric embeddings can lead to the same graph Laplacian, it is worthwhile noting that the eigenfunctions of the graph Laplacian also exhibit the spectral property. In Section \ref{sec:basis} we formally define the combinatorial Laplacian  as an operator over $\Omega$ considered as an oriented simplicial complex and show that the combinatorial Laplacian can be computed efficiently as a graph Laplacian on the dual graph of the simplicial complex. This definition also preserves the interpretation of the combinatorial Laplacian as a discretized version of $\nabla \cdot \nabla$. 

\section{Sliding Basis Optimization}
\label{sec:method}

In this paper, we  compactly
represent the material distribution as field parameterized by a weighted sum of the combinatorial Laplacian eigenfunctions computed over a volumetric mesh. We also reduce the number of 
design variables by truncating the eigenfunction expansion to bound the 
material field distributions considered in the optimization. This enables a 
significant speed up in the design optimization process. As opposed to 
previous applications using the Laplacian eigenfunction expansion, we do not 
use a fixed basis; instead we explore the basis space 
gradually from low frequency ones towards the higher frequencies. In 
addition to the computational benefits for optimization, this order reduction 
allows us to compute small number of eigenvectors as needed in contrast to  the
full eigenvalue decomposition which can be costly for large 
meshes~\cite{Song:2019multiscale}. In applying Laplacian basis to practical
material design problems, the need for enforcing material property bounds 
introduces large number of additional constraints to the optimization problem. 
We utilize logistic function based filters to avoid these extra 
constraints in both graded and discrete material design.

\subsection{Overview}

Given an input domain represented by a mesh $\Omega$ and a set of 
input goals (optimization objective and constraints), we optimize the (discretized) material 
field $\boldsymbol{\mathcal{F}}$ defined on $\Omega$ such that the input goals 
are satisfied. We parameterize $\boldsymbol{\mathcal{F}}$ as a weighted sum of well defined basis functions such that 
$\boldsymbol{\mathcal{F}} = \boldsymbol{B}\boldsymbol{w}$,  where 
$\boldsymbol{B}\in \mathbb{R}^{n_e  \times k}$ is a basis matrix whose  columns are the eigenvectors of the graph Laplacian (see Section \ref{sec:basis}), and 
$\boldsymbol{w} \in \mathbb{R}^k$ is the weight vector. Here, $k$ and $n_e$ 
correspond to number of selected basis functions and number of elements in 
$\Omega$. The approach of parameterizing fields over surfaces via  weighted Laplacian eigenfunctions is well known \cite{levy2006laplace}, and we observe that when solving inverse problems optimal fields may be defined by finding optimal weights  $\boldsymbol{w} $ for precomputed eigenfunction expansions.

 Due to the spectral property of the Laplacian eigenfunction basis, we start with the `low-frequency' basis functions whose support captures large portions of $\Omega$, and iteratively slide 
on the  ordered basis axis towards the `higher-frequency' basis functions whose support includes finer features, as 
the optimization progresses. This idea is inspired by the observations made in \cite{taubin1995signal,levy2006laplace} where analogies to signal processing, such as using low-pass filters in the frequency domain by projecting signals to the Fourier basis, can be applied to the Laplacian eigenfunctions for non-trivial geometric processing. In  geometric processing applications, the geometry  $\Omega$ is treated as the signal so that filters on the eigenfunction supports are used to perform editing operations such as smoothing the shape. For example, low-pass filters help remove fine features such as surface noise, sharp creases etc but preserve most of the shape's larger features. In this paper we consider the material field (defined over $\Omega$) as the signal that is supported by only as many eigenfunctions as required to satisfy performance objectives and constraints. The sliding basis algorithm provides the numerical framework to efficiently explore the ordered basis space  and construct a parameterization of the optimal material field.

\begin{figure}
\centering
\includegraphics[scale=0.2]{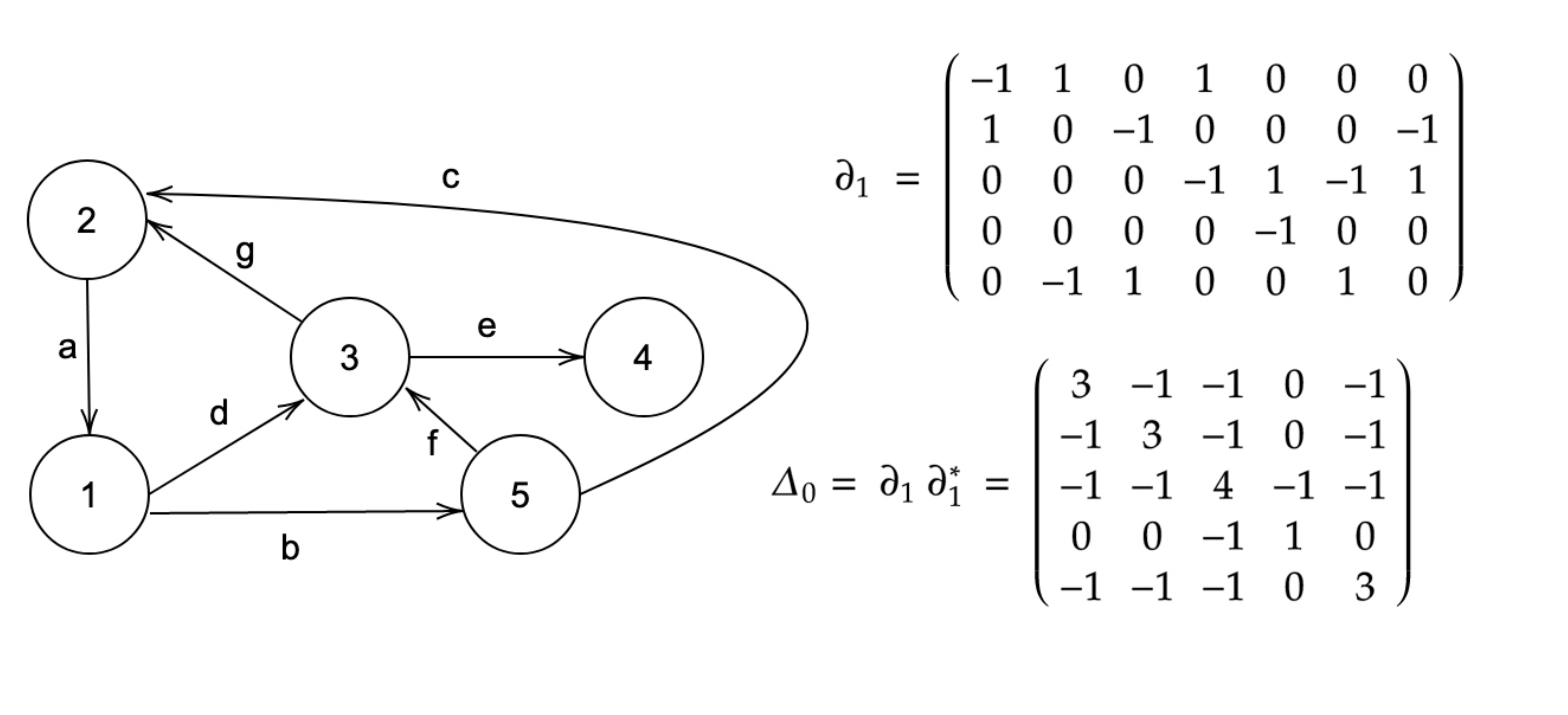}
\caption{A graph considered as an arbitrarily oriented 1-complex and the corresponding boundary operator $\partial_1$ is shown. Notice that $\partial_1 \partial_1^*$ is the traditional definition of the graph Laplacian. Here $^*$ represents matrix transpose.}
\label{fig:graph}
\end{figure}

Figure~\ref{fig:overview} illustrates our pipeline for an example design problem where the goal is to synthesize a material field for a solid rocket fuel, such that the burning fuel induces a prescribed thrust profile (thrust as a function of time). Such a problem exhibits the characteristics of where our approach is most applicable; the underlying domain  $\Omega$ is largely fixed and the goal is to synthesize a material field over $\Omega$, and the analysis for the thrust may be provided by a custom numerical procedure. We will describe this problem and its solution in greater detail in Section \ref{sec:applications}. In general, for each design scenario, we assume that the design goals can be described 
through an objective, $f$ and a set of constraints, $g_i$ in the form of a 
general optimization problem:

\begin{equation}
\begin{aligned}
& \underset{\boldsymbol{\boldsymbol{w}}}{\text{min}}
& &f(\boldsymbol{w}) \\
& \text{s.t.} & &  g_i(\boldsymbol{w}) \leq 0  \\
\end{aligned}
\label{eq:generalOpt}
\end{equation}

\noindent where the optimization is coupled with a physical analysis. Note that our model reduction method is 
differentiable. Therefore, if the analytical gradients are already derived for 
the full material field, $\frac{\partial f}{\partial \boldsymbol{\mathcal{F}}} 
$, gradients for the reduced order problem can be computed through a simple 
chain rule multiplication

\begin{equation}
\frac{\partial f}{\partial \boldsymbol{w}} = \frac{\partial f}{\partial 
\boldsymbol{\mathcal{F}}} \frac{\partial \boldsymbol{\mathcal{F}}}{\partial 
\boldsymbol{w}}
\label{eq:chainRuleAnalyticalGrads}
\end{equation}

\noindent where $\frac{\partial \boldsymbol{\mathcal{F}}}{\partial 
\boldsymbol{w}} $ is the constant reduced order basis matrix, $\boldsymbol{B} $.

\begin{figure}[]
  \centering  
   \includegraphics[trim = 0in 0in 0in 0in, clip, width = 
\columnwidth]{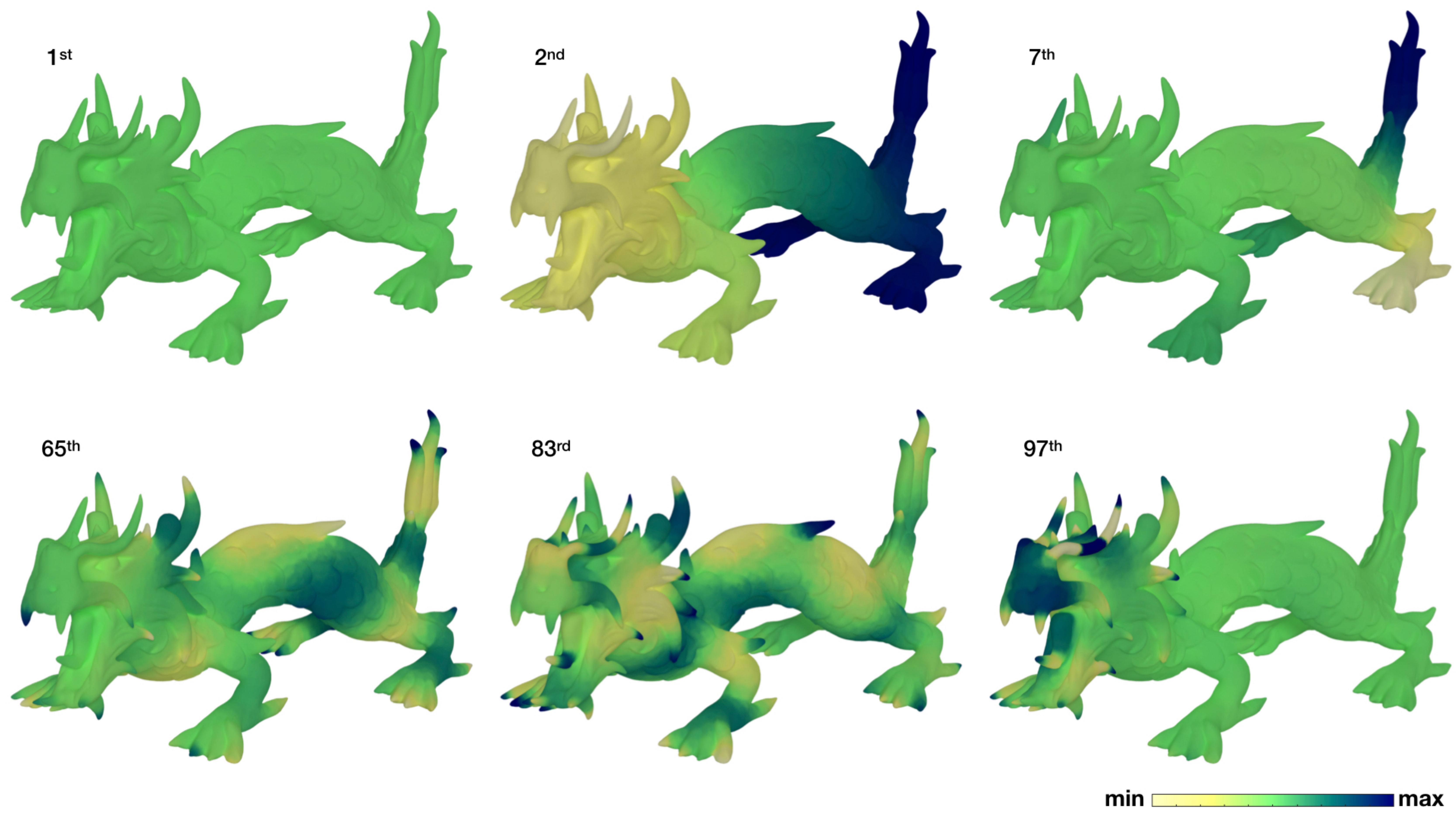}
  \caption{Spectral properties of Laplacian basis on an arbitrarily selected 
set. Notice the level of 
detail in the spatial field increases with the higher frequency basis 
functions.}
  \label{fig:spectralProps}
\end{figure}

\subsection{Laplacian Eigenfunction Basis}
\label{sec:basis}

The $i^{\mathrm{th}}$ combinatorial 
Laplacian operator over a finite oriented simplicial complex is defined as 
follows \cite{horak2013spectra, goldberg2002combinatorial, duval2002shifted}
\begin{equation}
\Delta_i = \partial_{i+1} \partial_{i+1}^* + \partial_i^* \partial_i
\end{equation}

Here $\partial_{i}$ represents the boundary operator for the $i^{\mathrm{th}}$ dimensional simplices,  and  $\partial_i^*$ is the adjoint operator of the boundary (aka the $i^{\mathrm{th}}$ co-boundary). The algebraic topological definition of the boundary operator on an oriented simplicial complex can be written as a matrix.
\begin{equation}
\partial_i = 
\begin{pmatrix}
\iota_{1,1} & \iota_{1,2} & \cdots & \iota_{1,n} \\
\iota_{2,1} & \iota_{2,2} & \cdots & \iota_{2,n} \\
\vdots  & \vdots  & \ddots & \vdots  \\
\iota_{m,1} & \iota_{m,2} & \cdots & \iota_{m,n} 
\end{pmatrix}
\end{equation}
Here the oriented simplicial complex $\Omega$ has $m$ $i-1$ dimensional faces and $n$ $i$ dimensional faces; $\iota_{p,q}$ is $1$ if the $p^{\mathrm{th}}$ $i-1$ dimensional simplex is a positively oriented face of the $q^{\mathrm{th}}$ $i$ dimensional simplex, and $-1$ if the $i-1^{\mathrm{th}}$ dimensional simplex is negatively oriented. Considering finite simple graphs as (arbitrarily) oriented simplicial complexes of dimension 1, we notice that $\Delta_0$ is the graph Laplacian (since $\partial_0 = 0$). A simple example is shown in Figure \ref{fig:graph}. Note that when we have an oriented simplicial complex, the boundary is a linear operator written as a matrix multiplication. But the adjoint operator avoids the need to compute the boundary this way due to the duality with the graph Laplacian.

Considering a tetrahedral mesh $\Omega$ as an oriented simplicial complex of dimension 3 and computing $\Delta_3$ we obtain the definition $\Delta_3 = \partial_3^* \partial_3$. Suppose we construct the dual complex $\bar{\Omega}$, where the $i$-simplices of the primal complex $\Omega$  are mapped to $3-i$ simplices in $\bar{\Omega}$. Then we observe the operator $\partial_3^* \partial_3$ over $\Omega$ is identical to $\partial_1 \partial_1^*$ over $\bar{\Omega}$. Thus $\Delta_3$ over $\Omega$ is the graph Laplacian over $\bar{\Omega}$. Therefore we define the combinatorial Laplacian $\Delta_3 = \boldsymbol{\mathcal{L}} = \boldsymbol{D} - 
\boldsymbol{A}$, where $\boldsymbol{D}$ is a diagonal matrix with each entry 
representing the element degree and $\boldsymbol{A}$ is the adjacency matrix 
given by

\begin{equation}
\boldsymbol{A}(i,j)=
\begin{cases}
1  \text{  \; if elements i and j share a face,} \\
0 \text{  \; otherwise}
\end{cases}
\label{eq:adjacencyMat}
\end{equation}

We also note that $\partial_1 \partial_1^*$ is the (combinatorial) divergence of 
the gradient on the dual complex. The Laplacian eigenfunction bases are then 
derived by solving for the eigenvectors of 
$\boldsymbol{\mathcal{L}}$

\begin{equation}
\lambda_i \boldsymbol{e_i} =  \boldsymbol{\mathcal{L}}  \boldsymbol{e_i},      
\forall i
\label{eq:eigenDecompLaplace}
\end{equation}

\noindent where $\lambda_i$ and $\boldsymbol{e_i}$ are the eigenvalues and 
eigenvectors of $\boldsymbol{\mathcal{L}}$, respectively. The basis matrix 
$\boldsymbol{B}$ can then be assembled by concatenating the eigenvectors side 
by side, $ \boldsymbol{B} = [\boldsymbol{e_1}, \boldsymbol{e_2}, ..., 
\boldsymbol{e_k}] \in \mathbb{R}^{n_e  \times k}$, where $n_e$ represents the 
number of 3-simplices (tetrahedral elements). Using weights of the basis 
functions $ \boldsymbol{w} = [w_1, w_2, ..., w_k]^T$ as design variables, we 
represent the material field as $\boldsymbol{\mathcal{F}} = 
\boldsymbol{B}\boldsymbol{w}$. Although it is possible to compute all the 
available basis functions \ie~$ k=n_e$, we avoid this costly operation in our 
sliding basis optimization by starting with a small number of functions $k \ll 
n_e$ and introduce additional eigenfunctions as needed by simply concatenating 
new basis vectors to the right side of the matrix B.  To compute a small subset 
of eigenvectors in Eq.~\eqref{eq:eigenDecompLaplace}, we utilize the Spectra 
library~\cite{spectra} in our implementation.

\begin{algorithm}
\small
\SetAlgoLined
\textbf{Input:}  $n_{opt}$, $n_{s}$, $s_{max}$ \\
\textbf{Output:}  Optimized basis weights, $\boldsymbol{w}$ \\
$i_{sb} \leftarrow 0$ \Comment{Index for the first active basis set}\\
$it_s \leftarrow 0$ \Comment{Sliding iteration} \\
$f \leftarrow 1/\epsilon$ \Comment{A large number} \\
$\boldsymbol{w} \leftarrow \varnothing $ \Comment{Optimized basis weights}\\
\While{not converged \textbf{or} $it_s < s_{max} $}{
 $\boldsymbol{w}_s$ $\leftarrow$  Initialize() \Comment{Weights for active 
basis functions} \\
 ($\boldsymbol{w}_s$, $f_s$) $\leftarrow$ Optimize($i_{sb}$, $n_{opt}$) \\

 \eIf{$f - f_s \geq \epsilon$}{  
       $\boldsymbol{w}    \leftarrow  [ \boldsymbol{w}[0 : i_{sb}],~ 
\boldsymbol{w}_s ]$\\
       $f \leftarrow f_s$\\
       $it_s \leftarrow 0$ \\
   }{
       $\boldsymbol{w}   \leftarrow [ \boldsymbol{w}, \boldsymbol{0}]$\\
       $it_s \leftarrow it_s + 1$ \\
  }
  
  $i_{sb} \leftarrow i_{sb} + n_{s}$ \\
 }
\caption{Sliding basis optimization }
\label{alg:slidingBasis}
\end{algorithm}

\subsection{Optimization}
\label{sec:algorithm}
\begin{figure}[]
  \centering  
  \includegraphics[trim = 0in 0in 0in 0in, clip, width = 
\columnwidth]{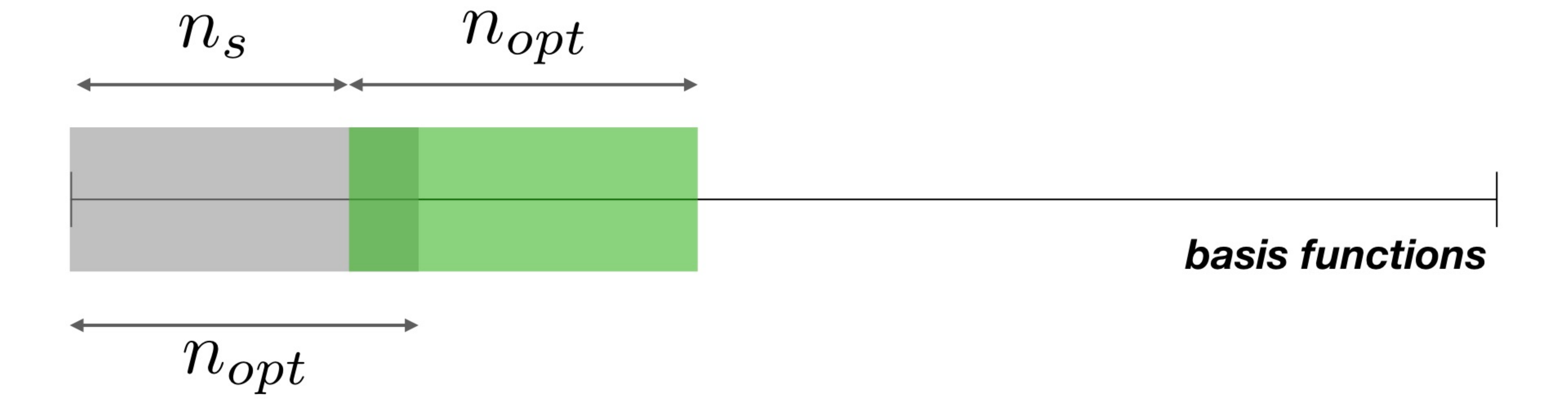}
  \caption{Sliding basis optimization starts by optimizing for the first 
$n_{opt}$ bases. Then, the selected basis are shifted by $n_s$ and the 
optimization is performed for the new set of $n_{opt}$ bases. This slide and 
optimize procedure continues until convergence.}
  \label{fig:slidingSchematic}
\end{figure}

We exploit the spectral property of the Laplacian 
basis~(Figure~\ref{fig:spectralProps}) to explore the design space and 
iteratively optimize the material field. Figure~\ref{fig:slidingSchematic} 
illustrates the basic 
idea of our sliding basis optimization. Here, $n_{opt}$ and $n_s$ are the 
number of active basis functions (\ie optimization variables) and the amount of 
sliding at each sliding basis optimization step, respectively. We explore the 
basis space through multiple optimization operations where only the weights 
corresponding to the 
active $n_{opt}$ bases are optimized at a time. We optimize for only a small 
number of bases and we slide on the ordered basis axis by $n_s$ bases and 
perform another optimization with a new set of $n_{opt}$ variables. The sliding 
iterations continue until convergence. 

Algorithm~\ref{alg:slidingBasis} describes the approach in detail. Given 
$n_{opt}$, $n_s$ and $s_{max}$ (the number of maximum trials before stopping if 
there is no significant improvement in objective value), the algorithm returns 
a set $\boldsymbol{w}$ of
optimized basis weights. We choose $n_s < 
n_{opt}$ so that there is an overlapping set of active basis function 
whose weights are re-optimized 
(Figure~\ref{fig:slidingSchematic}) over the previous step.  At 
each sliding basis optimization step, the weights $\boldsymbol{w}_s$ corresponding to the active 
basis functions (initially the first $n_{opt}$ 
eigenvectors of $\mathcal{L}$), are optimized to achieve 
the lowest possible objective value, $f_s$. Note that the optimization step 
here can be implemented using a commodity optimizer including  a gradient 
based or a stochastic one. In our examples, we use sequential quadratic 
programming (SQP)~\cite{Nocedal:2006} as it is an effective nonlinear 
programming method for general optimization problems. At each optimization 
step, we want 
to avoid getting 
stuck at the local minimum found in the previous step. This can happen, 
for example, if we initialize the weights for each new added 
basis to $0$ and use the previously optimized values for the weights of the 
overlapping 
basis functions.  In contrast, 
initializing all the weights of an active basis set to be $0$ or 
random-valued perturbs the initial condition away from the previous local 
minumum. If the resulting optimized weights yield an objective value $f_s$ 
that is lower than $f$, the weights are accepted and 
$\boldsymbol{w}$  is expanded to include $\boldsymbol{w}_s 
$.  If the objective 
value is not improved in the current iteration, the weights of the overlapping 
region are not modified and $\boldsymbol{w}$ is concatenated with $0$ weights 
corresponding to the newly added $n_s$ basis functions. Subsequently, the 
active set is 
modified by sliding towards the higher frequency basis functions. The sliding 
basis optimizations stops if the addition of the new basis does not 
significantly improve the objective or the maximum number of iterations are 
reached.

Suppose $k$ Laplacian basis functions are selected 
and a gradient based optimization approach such as SQP is utilized to solve 
a material design problem. The costliest step in such an 
approach is often the Hessian computation where the computational cost 
increases quadratically with the increasing number of design variables (\ie 
number of basis functions in our case), $\mathcal{O}(k^2)$. In cases involving 
black-box analysis of the objective (e.g. when the implementation for the 
physical simulation to evaluate the objective is unavailable), the conventional 
approach of optimizing all $k$ design variables at once results in $k^2$ 
analysis runs at
each optimization step to construct the Hessian matrix. Given that analysis/ 
physical simulation 
is often expensive, the quadratic relationship makes the use of 
large $k$ impractical by creating computational bottlenecks. In the sliding 
basis 
optimization approach, the total computational cost is kept 
lower by exploring the same $k$ basis functions gradually, i.e. $n_{opt}$ 
functions at a time. In 
this case, only $n_{opt}^2$  analysis runs are required to construct the 
Hessian matrix. Here, it is important to note that $n_{opt} << k$. As the 
optimizer needs to be reinitialized after each sliding iteration in our 
approach, $p = (k - 
n_{opt}) / n_s + 1$ complete optimization operations are performed to cover $k$ 
basis functions. Assuming same number of iterations are performed in each 
optimization operation and $n_s \rightarrow n_{opt}$, this translates to 
reducing the total computational cost by a factor of up to $n_{opt}/k$ over 
optimizing for fixed $k$ basis.

\begin{figure}
  \centering  
  \includegraphics[trim = 0in 0in 0in 0in, clip, width = 
\columnwidth]{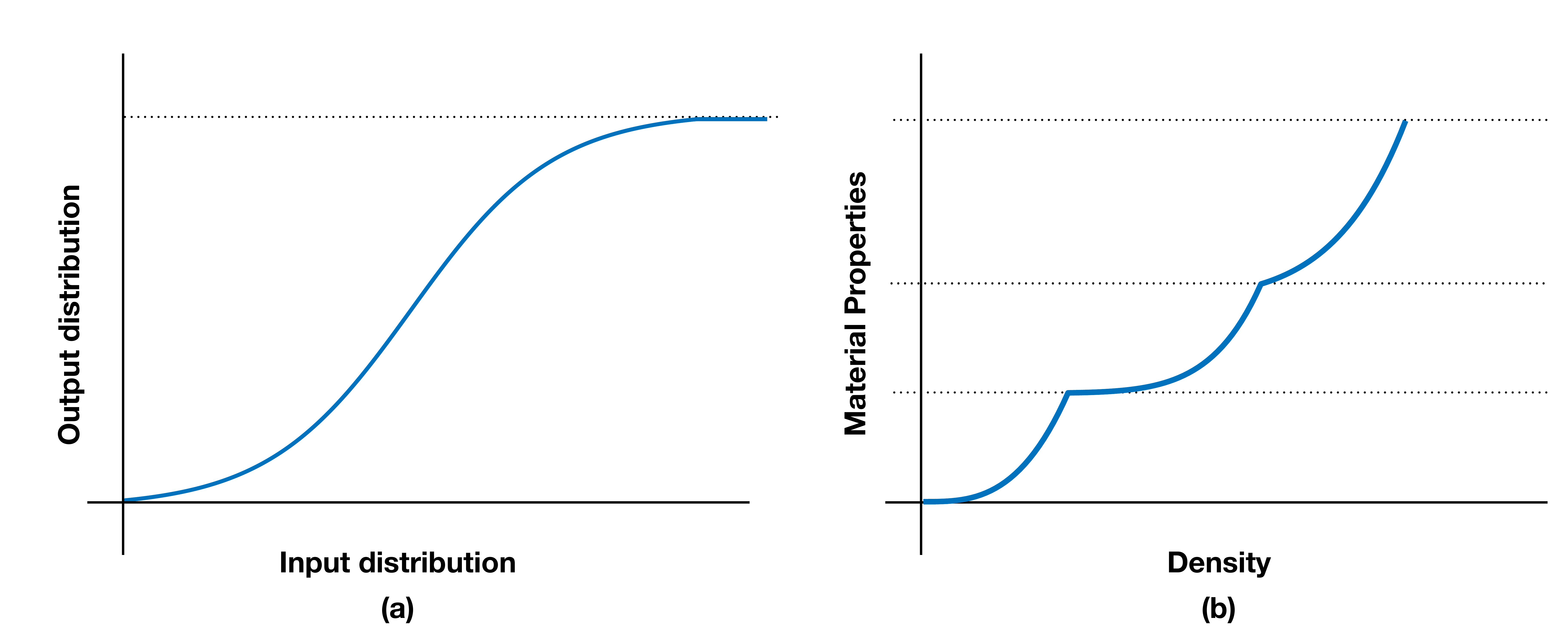}
  \caption{We utilize a gentle slope logistic function to bound the material 
distributions~(a) and ordered SIMP approach to achieve  discrete multi-material 
distribution~(b).}
  \label{fig:sCurve}
\end{figure}

For graded material design problems, bounds of the allowable material 
properties need to be enforced so that the optimized field can be manufactured. 
To enforce the bounds, one approach is to add additional linear inequality 
constraints in the form of $ \boldsymbol{B} \boldsymbol{w} \leq 
u_{mb} $ and $ \boldsymbol{B}  \boldsymbol{w} \geq 
l_{mb} $ to the general optimization problem given in 
Eq.~\eqref{eq:generalOpt}. However, this approach increases the number of 
constraint by 2*$n_e$ which could be in the order of hundred thousands for dense 
material distributions. This increase in the number of constraints slows down 
the optimization process significantly, especially for cases where the 
analytical gradients are not available. Instead, we use a filtering approach to 
bound the material distribution of the field without introducing additional 
constraints. We use a logistic function

\begin{equation}
 l(x) =  l_{mb} + \frac{u_{mb}-l_{mb}}{1+exp(-\kappa(x))}
\label{eq:logisticFunction}
\end{equation}

\noindent where $\kappa$ is the steepness parameter set to give a gentle slope 
as shown in Figure~\ref{fig:sCurve}.a. After the material field is computed as 
a weighted combination of the basis functions, we utilize the logistic function 
to enforce the bounds of the manufacturing technique. This approach provides a 
differentiable way to limit the material properties for manufacturability. 
Similarly, we enforce material constraints for multi-material optimization of 
discrete sets by combining this filtering approach with the penalization 
methods (Figure~\ref{fig:sCurve}.b). We will now discuss these properties in the context of non-trivial 
design applications.

\section{Applications}
\label{sec:applications}

\subsection{Graded Solid Rocket Fuel Design}

\begin{wrapfigure}{r}{0.35\columnwidth}
  \begin{center}
    \includegraphics[width=0.35\columnwidth]{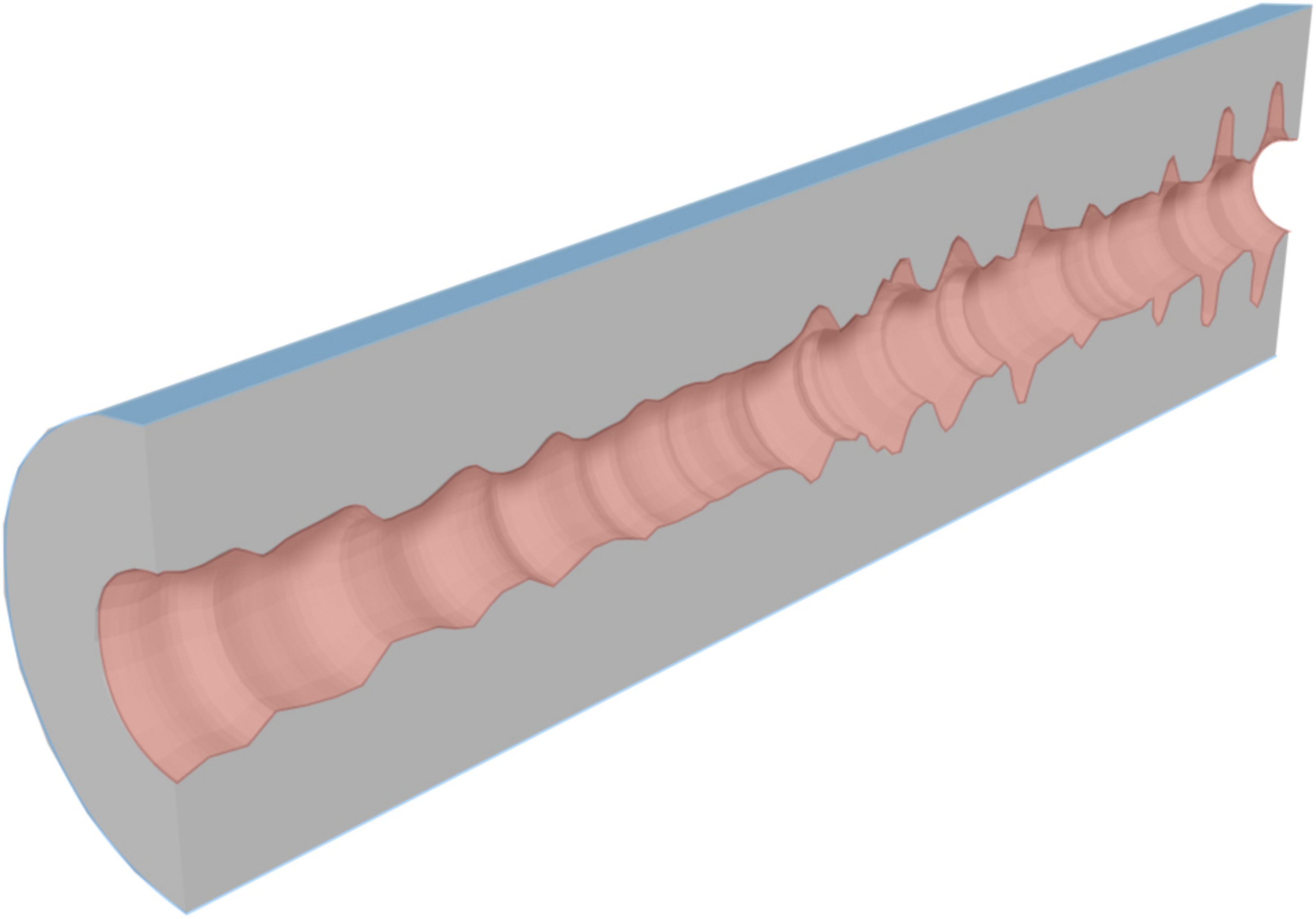}
  \end{center}
\end{wrapfigure}
In this application, our goal is to develop a computational tool to design a 
multi-material solid rocket propellant. Solid rocket propellant is shown at right with the inner surface geometry (where ignition takes place) in red and outer casing in blue.The 
propellant design should satisfy two main requirements:
(1) The thrust generated as the propellant burns needs to match a given target 
thrust profile (2) No insulation should be required at the outer casing. 
The former requirement ensures that the rocket behaves as desired during its 
use. The latter requirement indicates that at the moment prior to burn out, all 
the casing surface is covered with some propellant material that will vanish at the same 
time, ensuring no part of the casing surface is exposed to on-going 
burning. Elimination of the insulation is an important innovation to 
substantially reduce the rocket weight. The time-varying nature of this problem 
brings up extra computational complexity in the analysis. Our model reduction 
and efficient sliding basis exploration plays a critical role for the solution 
of such a problem.

\paragraph{Physical Analysis}
We first examine the physical relationships between a provided thrust 
profile (required thrust vs time) and the burn rate distribution of the rocket 
propellant and develop a 
boundary value problem to simulate the solid propellant burn. At any time $t$, 
the relationship between the thrust profile, $th(t)$ and the 
pressure inside the combustion chamber, $P_c(t)$ is given as

\begin{equation}
P_{c}\left(t\right)=\frac{th\left(t\right)}{C_{f}A_{t}},
\end{equation}

\noindent where $C_{f}$ and $A_{t}$ are thrust coefficient and throat area, 
respectively. The mass flow rate of the exhaust gas 
emanating from the rocket nozzle is proportional to the pressure inside the 
chamber,

\begin{equation}
\dot{m}_{out}\left(t\right)=\frac{A_{t}P_{c}\left(t\right)}{c_{s}}.
\end{equation}
Here $c_s$ represents the speed of sound. The mass flow rate emanating from the 
burn front can be calculated as a surface 
integral

\begin{equation}
\dot{m}_{in}\left(t\right)=\oiint_{\Gamma\left(t\right)}\rho_{p}\dot{r}\left(x,y
,z\right)d\sigma,
\end{equation}

\noindent where $\rho_{p}$ is the propellant density, 
$\dot{r}\left(x,y,z\right)$ is the spatial propellant burn rate distribution 
(as a function of the material at $(x,y,z)$) and $\Gamma\left(t\right)$ is the 
burn surface. The burn rate is related to the 
chamber pressure and the reference burn rate through the power law 

\begin{equation}
\dot{r}\left(x,y,z\right)=\dot{r}_{ref}\left(x,y,z\right)\left(\frac{P_{c}}{P_{
ref}}\right)^{n},
\end{equation}
where $\dot{r}_{ref} \left(x,y,z\right)$ is the spatial distribution of the 
reference burn rate, $P_{ref}$ is the reference pressure and $n$ is the kinetic 
constant. Assuming a simplified mass conservation 
$\dot{m}_{in}\left(t\right)=\dot{m}_{out}\left(t\right)$, the 
thrust profile can be derived as a function of the reference burn rate:

\begin{equation}
th\left(t\right) = \left( P_{ref}^{-n} I_{sp}^{1-n} A_t^{-n} c_s^n 
\oiint_{\Gamma\left(t\right)} \rho_p \dot{r}_{ref}\left(x,y,z\right)d\sigma 
\right)^{\frac{1}{1-n}}\label{eq:thrust}
\end{equation}

\noindent To close the Eq. \ref{eq:thrust}, we model the evolution of the burn 
surface $\Gamma\left(t\right)$
using the Eikonal equation

\begin{equation}
\left|\nabla\phi\right|=\frac{1}{\dot{r}_{ref}\left(x,y,z\right)}.\label{
eq:Eikonal}
\end{equation}

\noindent The level sets of $\phi$ define the surface of the burn front at 
time instance $t$,

\begin{equation}
\Gamma(t) = \left\{ \{x,y,z\} \in \mathbb{R}^3 | \phi\left(x,y,z\right)=t 
\right\}.\label{eq:level-set}
\end{equation}

We solve equations  \eqref{eq:thrust}-\eqref{eq:level-set} assuming an axial 
symmetry inside a cylindrical rocket chamber with length, $L$  inner radius, 
$r_{in}$ and outer radius, $r_{out}$. When solving the boundary value problem 
we ensure the burn surface at the final time step is equal to the outer rocket 
case, $\Gamma_{case}$ \ie $\Gamma\left(t_{end}\right)=\Gamma_{case}$.

\paragraph{Optimization Problem}

We minimize the $\boldsymbol{\ell}^2$ norm of the error in matching the 
thrust profile while constraining the inner burn surface to avoid placing any 
insulating material

\begin{equation}
\begin{aligned}
& \underset{\boldsymbol{w}}{\text{min}}
& & \sum_t(th(\boldsymbol{w}) - th_{target})^2 \\
& \text{s.t.} & & r_b(\boldsymbol{w})^i > r_{in}
\end{aligned}
\label{eq:optRocket}
\end{equation}

\noindent where $th_{target}$ and $th$ represent the target thrust profile and 
the current thrust profile achieved with the distribution 
$\boldsymbol{\mathcal{F}} = \dot{r}_{ref}(x,y,z)$ computed using the weights 
$\boldsymbol{w}$. Since the physical analysis solves a boundary value problem 
and constructs the level set of the burn surfaces starting from the outer case, 
we are able to optimize the material distribution such that the initial burn 
surface 
comes as a byproduct. For each material distribution, we represent the inner 
burn surface through a set of radius values, $r_b^i $, and check if they are 
all inside the allowable inner surface region defined by the radius value, 
$r_{in}$.

\subsection{Multi-Material Topology Optimization}

We apply our sliding basis optimization approach to the design of 
multi-material 
distributions with given discrete set of predefined materials for structural 
mechanics problems. 

\paragraph{Physical Analysis}

Assuming linear isotropic materials and small 
deformations, we solve the linear elasticity problem 
$\boldsymbol{K}\boldsymbol{u} = \boldsymbol{F}$ where $\boldsymbol{K}$, 
$\boldsymbol{u}$ and $\boldsymbol{F}$ are the stiffness matrix, nodal 
displacement vector, and nodal external force vector, respectively. In our 
implementation, we discretize the domain using tetrahedral elements 
characterized by linear shape functions assuming static load and fixed 
displacement boundary conditions.

\begin{table*}[t]
\caption{Performance of our sliding basis optimization algorithm for the graded 
solid rocket design problem on a variety of target thrust profiles. Performance 
of fixed basis reduced order optimization is also provided for comparison. Note 
that fixed basis optimization and sliding basis optimization cover the same 
basis functions.}
\begin{center}
\label{tab:statistics}
\begin{tabular}{@{\extracolsep{4pt}}l c c c c c c c c@{}}  
& & &&& \multicolumn{2}{c}{Fixed Basis} & \multicolumn{2}{c}{Sliding Basis} \\ 
\cline{6-7}  \cline{8-9}
Thrust Profile                 &  $n_{opt}$ & $n_s$ & $n_{slides}$ & Total 
Basis & Time & Objective/Error & Time& Objective/Error \\
\hline
Constant Acceleration     & 20&15&14&230 & 1178s & 349k/2.3\% & 288s & 
86k/1.1\%\\
Constant Deceleration     & 50&40&7&320 & 4896s & 867k/3.4\% & 621s & 
452k/2.7\%\\
Two Step                           & 20&15&7&125 & 191s & 102k/1.1\% & 69s & 
217k/1.4\%\\
Bucket                               & 20&15&24&380 & 1006s & 272k/1.8\% & 596s 
& 272k/1.8\%\\
\hline
\end{tabular}
\end{center}
\end{table*}

\paragraph{Optimization Problem}

We formulate the multi-material design 
optimization as a density based topology optimization problem with compliance 
minimization and mass fraction constraint as

\begin{equation}
\begin{aligned}
& \underset{\boldsymbol{w}}{\text{min}}
& &  \boldsymbol{u}^T  \boldsymbol{K (\boldsymbol{w})}  \boldsymbol{u} \\
& \text{s.t.} & & m(\boldsymbol{w})/m_0 \leq m_{frac} \\
& & & \boldsymbol{K(\boldsymbol{w})}\boldsymbol{u} = \boldsymbol{F}\\
\end{aligned}
\label{eq:topOpt}
\end{equation}

\noindent where $m$, $m_0$ and $ m_{frac}$ are mass of the current design, mass 
of the design domain fully filled with maximum density and prescribed mass 
fraction. Here, $\boldsymbol{u}^T  \boldsymbol{K}  \boldsymbol{u}$ represents 
the compliance of the structure. We adopt the ordered multi-material SIMP 
interpolation approach~\cite{Zuo:2017multi} since it does not introduce 
additional variables and computational complexity as the number of materials 
increase. We incorporate the interpolation step after computing the density 
field with the weights and basis functions and using the bounding filter to 
keep density values in $[0,1]$ limits as explained in Section 
\ref{sec:algorithm}. Additionally, we implemented the density filtering 
approach 
described in~\cite{Andreassen:2011top88}. This filter is often utilized to 
avoid checkerboard issues in traditional topology optimization. We observe 
similar issues as we use higher order basis functions although we use a reduced 
order approach. We found the density filtering helpful in avoiding those issues 
in our reduced order approach as well.

\begin{figure}
  \centering  
  \includegraphics[trim = 0in 0in 0in 0in, clip, width = 
\columnwidth]{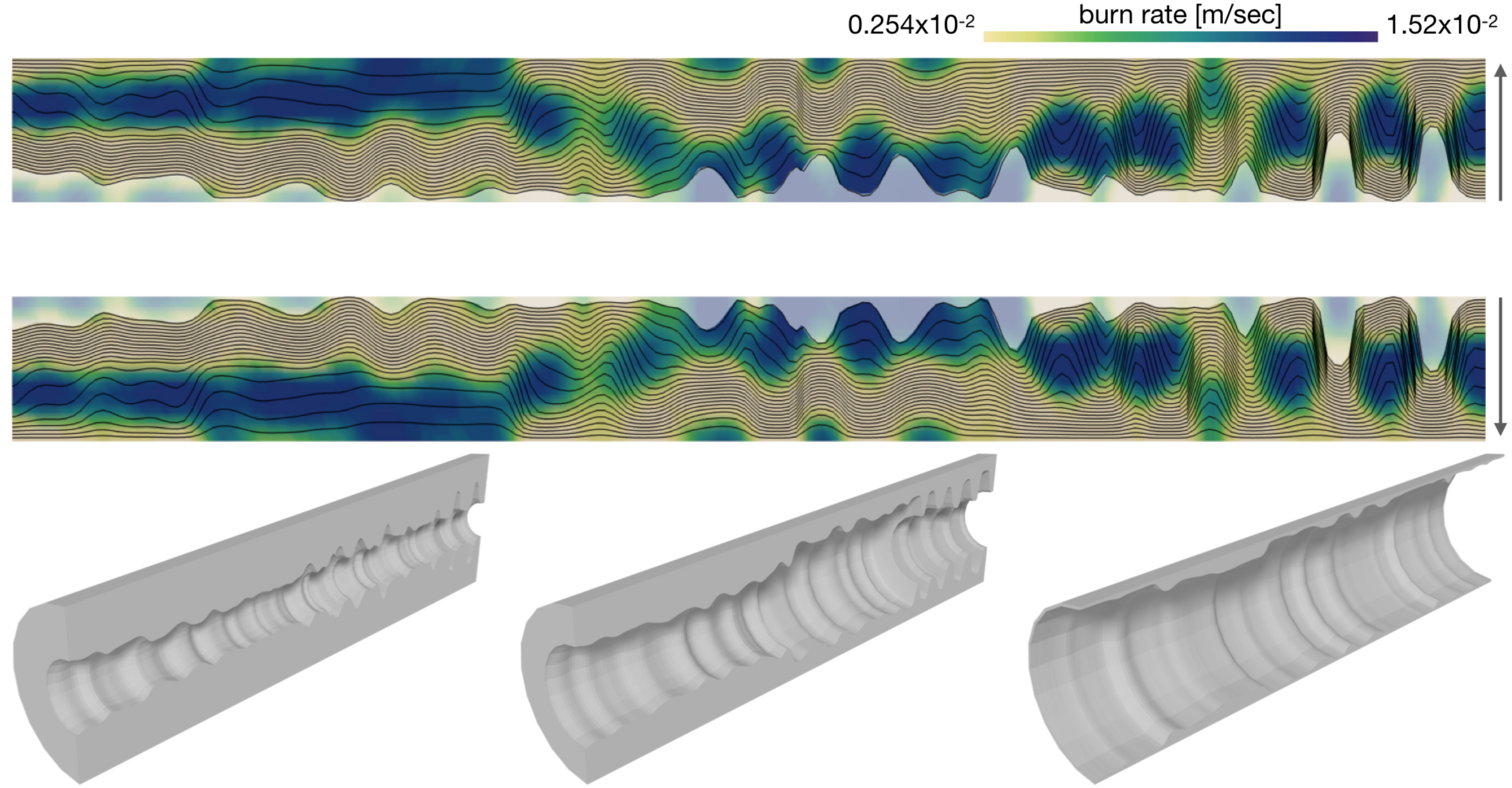}
  \caption{Top image shows the progression of the burn front~(black lines) on 
the cross section of solid rocket with the graded burn rate distribution for 
the constant acceleration thrust profile. The arrows on the right denote the 
burn direction. Notice the lines spread apart in high burn rate material 
regions~(blue) compared to the low burn rate regions(yellow) indicating faster 
burn. Bottom three cutouts show the solid rocket fuel at three stages of the 
burn propagation.}
  \label{fig:burnPropogation}
\end{figure}

\section{Results and Discussions}

We demonstrate the results of using sliding basis optimization in the 
applications described in Section \ref{sec:applications}. We discuss the 
progression of the optimization during sliding steps, 
performance gain and effect of sliding amount.

\begin{figure*}
  \centering  
  \includegraphics[trim = 0in 0in 0in 0in, clip, width = 
\textwidth]{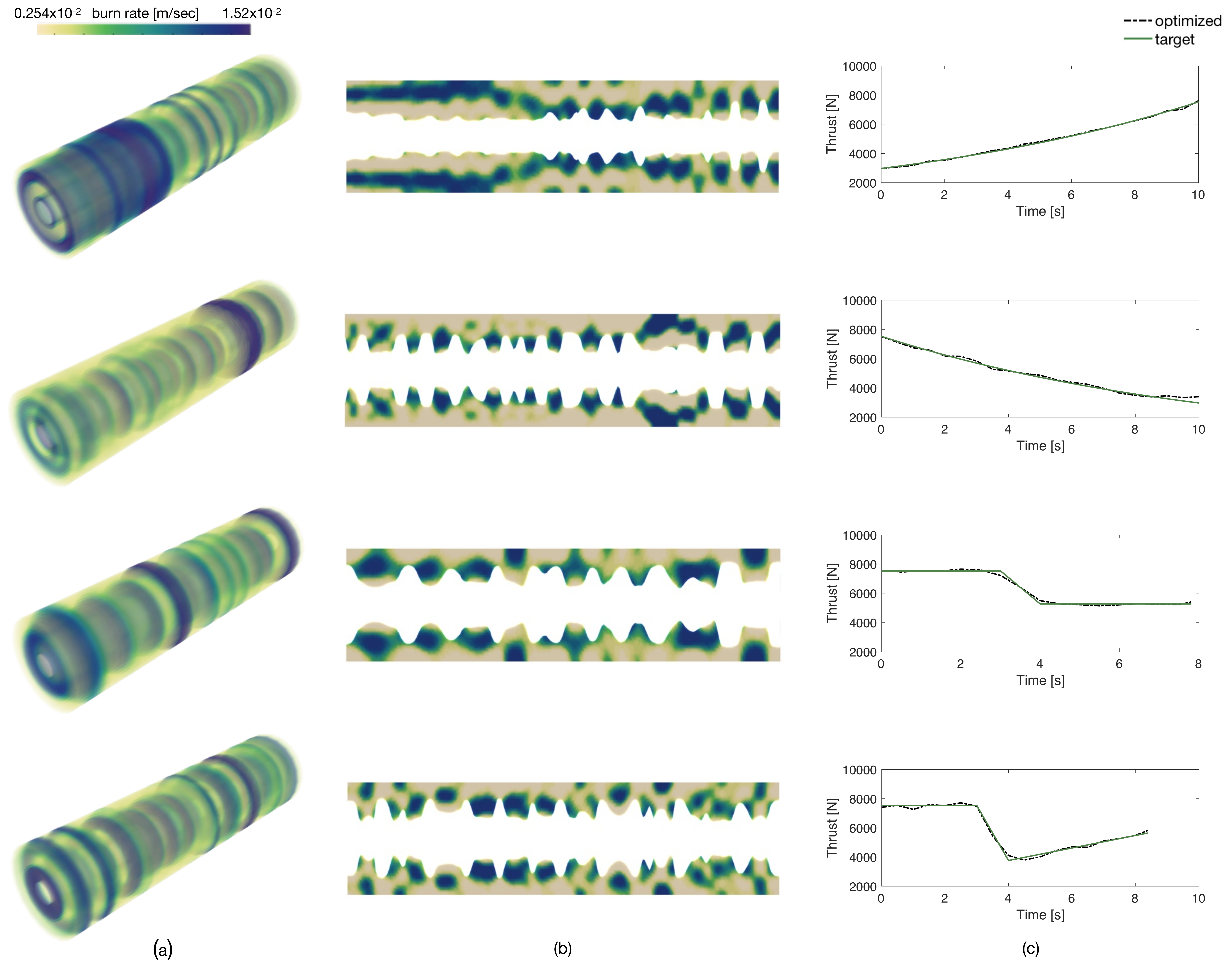}
  \caption{Solid rocket fuel design results with volume render of the burn rate 
distribution~(a), cross section of the rocket geometry with inner surface~(b) 
and thrust profile plot~(c). The images and plots from top to bottom represent 
the results of constant acceleration, constant deceleration, two step and 
bucket profiles.}
  \label{fig:rocketResults}
\end{figure*}

\paragraph{Graded solid rocket fuel design}The sliding basis optimization 
results for four different thrust profiles (constant acceleration, constant 
deceleration, two step and bucket) are presented in 
Fig.~\ref{fig:rocketResults}. Since our physical analysis assumes axially 
symmetric material distributions, we parameterize the cross section and compute 
the basis functions on it. We treat the analysis (solving the boundary value 
problem described in Section \ref{sec:applications}) as a black-box solver and 
do not 
derive the analytical gradients for this application to show the effectiveness 
of our sliding basis optimization approach. As shown in 
Fig.~\ref{fig:burnPropogation}, we parameterize the burn rate distribution on 
the whole rectangular cross section. For all of our examples, we used 3000 quad 
elements on the cross section. Due to the boundary value problem formulation 
which takes the last burn front~(outer cylindrical surface) as input, the inner 
burn surface is computed as a byproduct of the simulation. After the 
optimization is completed, we mask out the portions of the cross section that 
are beyond the inner surface since these portions are not needed to achieve the 
desired target thrust profile behavior.

One challenge in matching the thrust profiles is to be able to reduce the 
thrust significantly through the end of the burn process~(\eg constant 
deceleration profile). At a given time, the thrust is proportional to the area 
of 
the burn front surface. Since the surface area naturally increases as the burn 
surface propagates from inside to the outside of the cylinder, reducing thrust 
requires complex material distributions that can reverse this natural tendency 
to increase burn surface area. Therefore, the constant deceleration profile is 
more challenging and requires more complex material distributions than the 
constant acceleration profile. Table~\ref{tab:statistics} reports the parameters 
used in our examples and performance of our sliding basis optimization through time, objective value and the average percentage error between the target and optimized thrust profiles. While we 
use 20 optimization variables, $n_{opt}$ for the constant acceleration, two step 
and bucket profiles, we observe that using 50 optimization variables results in 
better performance for constant deceleration profile. We believe this is mainly 
due to the challenging nature of this profile that requires more complex 
distributions that can be achieved using more basis functions.

\begin{figure*}[]
  \centering  
  \includegraphics[trim = 0in 0in 0in 0in, clip, width = 
\textwidth]{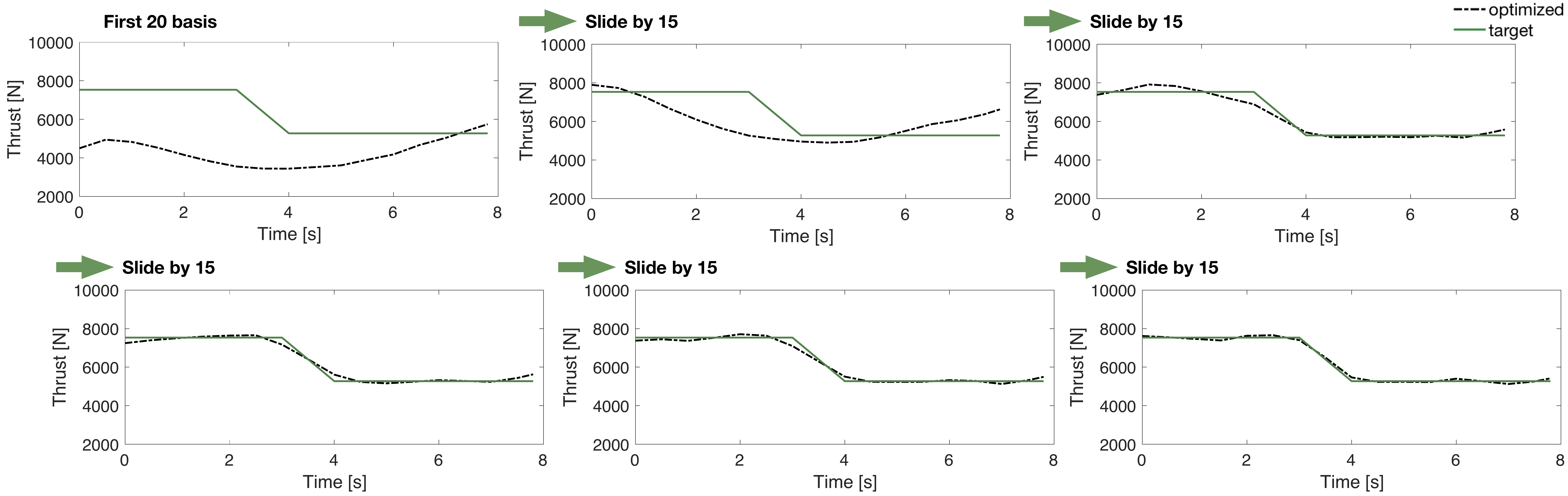}
  \caption{Progression of thrust profile match through the sliding basis 
optimization. With only first few basis functions, the results may be far from 
the desired. However, the optimizer recovers from those results with the 
additional sliding optimization steps since more basis functions are used.}
  \label{fig:slidingOptProgression}
\end{figure*}

Figure~\ref{fig:slidingOptProgression} shows the target and optimized 
thrust profile results during six steps of sliding basis optimization for two 
step thrust profile. The first optimization step that uses only twenty basis 
functions does not match the target profile well since the small number of 
basis functions is not enough to create complex enough material distributions 
for this case. In the consecutive steps, however, the resulting thrust profiles 
progressively match the target better as more basis functions are incorporated 
with each slide. Finally, the sliding optimization stops when the convergence 
criteria is satisfied. For this application, we used 5\% error in the profile 
match as the convergence criteria.

We compare the sliding basis optimization to the reduced order approach using 
a predetermined number of Laplacian basis functions during the optimization 
such that all basis functions are optimized 
simultaneously. Note that this `fixed basis' approach is already a reduced 
order method and significantly improves optimization performance by projecting 
optimization variables into lower dimensional space. Table~\ref{tab:statistics} 
reports the performance of both fixed and sliding basis optimization methods. 
Our sliding basis approach can speed up the optimization process up to 8 times 
over the fixed basis method while exploring the same number of basis functions. 
In addition to the time gain, sliding basis optimization results in better 
objective minimization performance in most of our examples. We believe this may 
be due to the random initialization at each sliding step that acts as local 
perturbations and alleviate local minima issues of general nonlinear 
optimization problems. All computations are performed and recorded on a 
computer with 16GB memory and 3.1GHz i7 processor.

\begin{figure}
  \centering  
  \includegraphics[trim = 0in 0in 0in 0in, clip, width = 
\columnwidth]{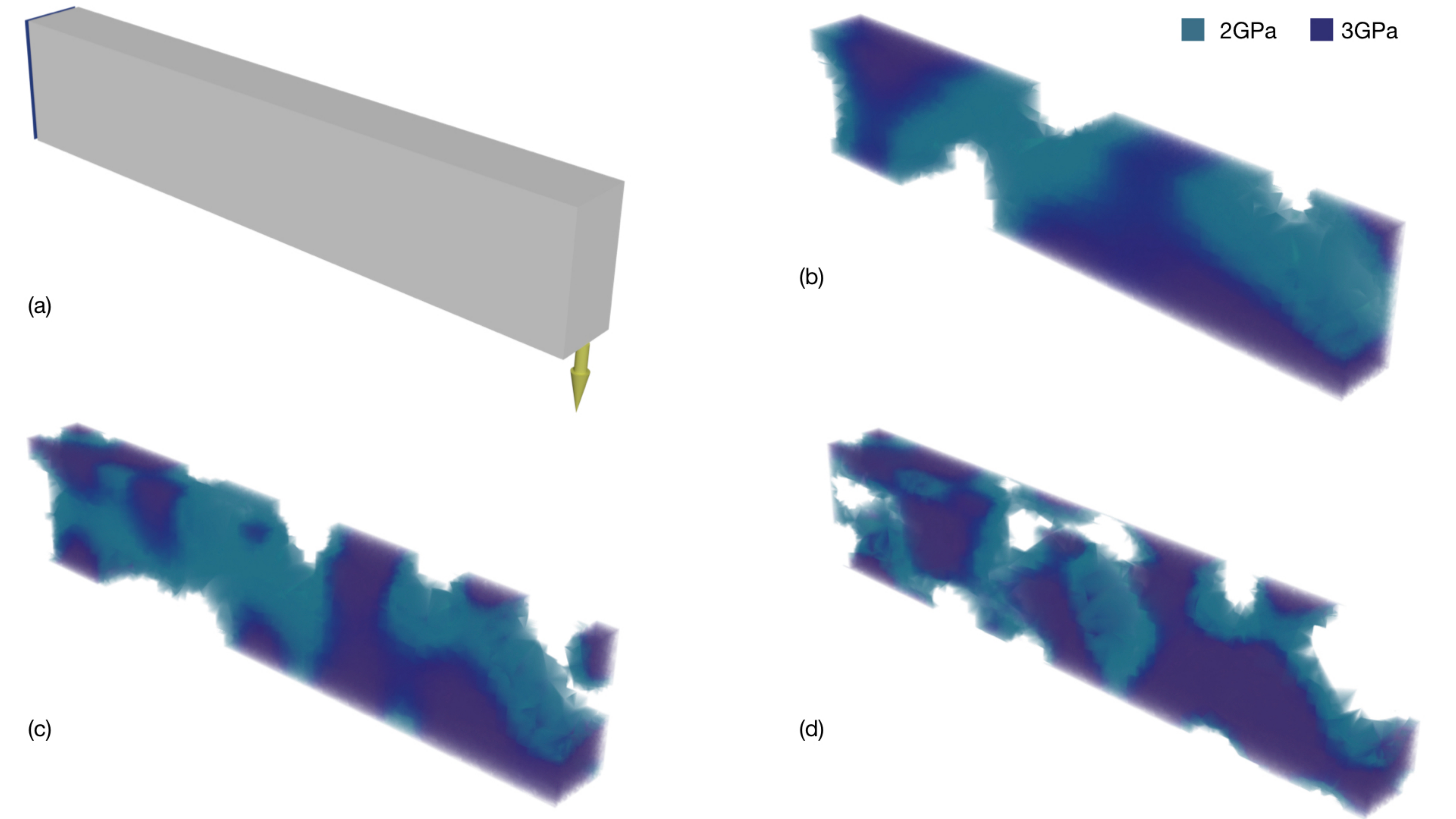}
  \caption{Multi-material topology optimization results for the cantilever beam 
problem throughout the sliding steps. (a) problem setup with boundary 
conditions~(blue regions) and external loads~(yellow arrows). (b)-(d): 
optimized material distributions for two materials and void. }
  \label{fig:topOptBeam}
\end{figure}

\paragraph{Multi-material topology optimization}

Figure~\ref{fig:topOptBeam} presents the problem setup (a) and the optimized 
material distributions for three steps of the sliding basis optimization (b-d) 
for the cantilever beam. For this beam example, we use a discrete set of void 
and two materials with normalized density values of $0,~0.1$ and $1$ and 
Young's modulus values of $0,~2GPa$ and $3GPa$. From 
Figure~\ref{fig:topOptBeam}(b) to Figure~\ref{fig:topOptBeam}(d), number of 
optimized basis functions increase from 20 to 170 indicating the complexity of 
material distribution increases as the number of basis functions increases. This 
sliding basis optimization is performed using $n_{opt} = 20$ and $n_s = 15$ with 
mass fraction constraint, $m_{frac}$ of 0.5 on a mesh with 19567 tetrahedral 
elements. Using numerical gradients (\ie treating simulation as black-box), we 
observe that sliding basis optimization results in approximately 3 times faster 
computation time ($87mins$ vs $259 mins$) compared to fixed basis optimization 
with 170 basis.

\begin{figure}
  \centering  
  \includegraphics[trim = 0in 0in 0in 0in, clip, width = 
\columnwidth]{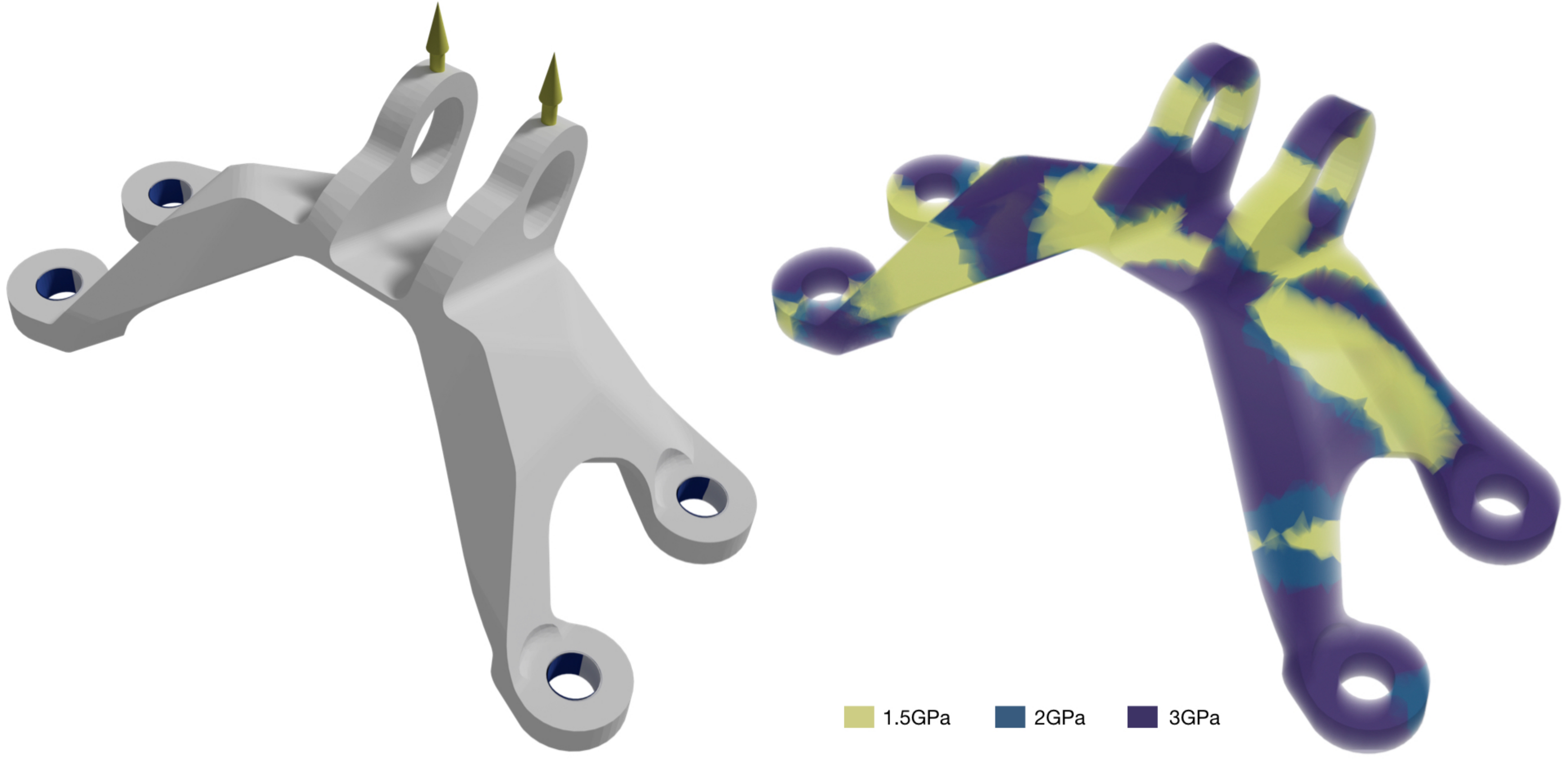}
  \caption{Multi-material topology optimization result for the bracket model. 
Left: problem setup with boundary conditions~(blue regions) and external 
loads~(yellow arrows). Right: optimized material distribution is given for  
three material set.}
  \label{fig:topOptBracket}
\end{figure}

The optimized material distribution of the bracket model is given in 
Figure~\ref{fig:topOptBracket} for a discrete set with three 
materials~(normalized densities: $0.1,~0.3,~1$ and Young's modulus: 
$1.5GPa,~2.5GPa,~3GPa$). It can be observed that the optimizer places the 
strongest material on the load paths an around the high stress regions such as 
where the boundary conditions and forces are applied. The optimization is 
performed using $n_{opt} = 20$ and $n_s = 15$ with mass fraction constraint, 
$m_{frac}$ of 0.5 on a mesh with 174454 tetrahedral elements. For this 
particular example, instead of using the numerical gradients, we derived and 
utilized the analytical gradients. The analytical gradient computation is an 
extremely fast operation for compliance minimization. Thus, the only costly 
operation during the optimization is the linear solve while finding the 
Lagrange multipliers. However, the computational cost of a linear solve 
operation with a matrix of size 20 or 200 are marginally different for state of 
the art linear solvers. Therefore, we do not observe a computational gain for 
using the sliding basis optimization over the fixed basis optimization for this 
particular compliance minimization problem when analytical gradients are 
utilized. However, note that fixed basis optimization already reduces the 
computational cost of the linear solve from a matrix of size 174454 to a matrix 
of 200. Furthermore, sliding basis optimization could provide more significant 
computational gain over fixed basis for problems with costly analytical gradient 
computations.

\begin{figure}[]
  \centering  
  \includegraphics[trim = 0in 0in 0in 0in, clip, width = 
\columnwidth]{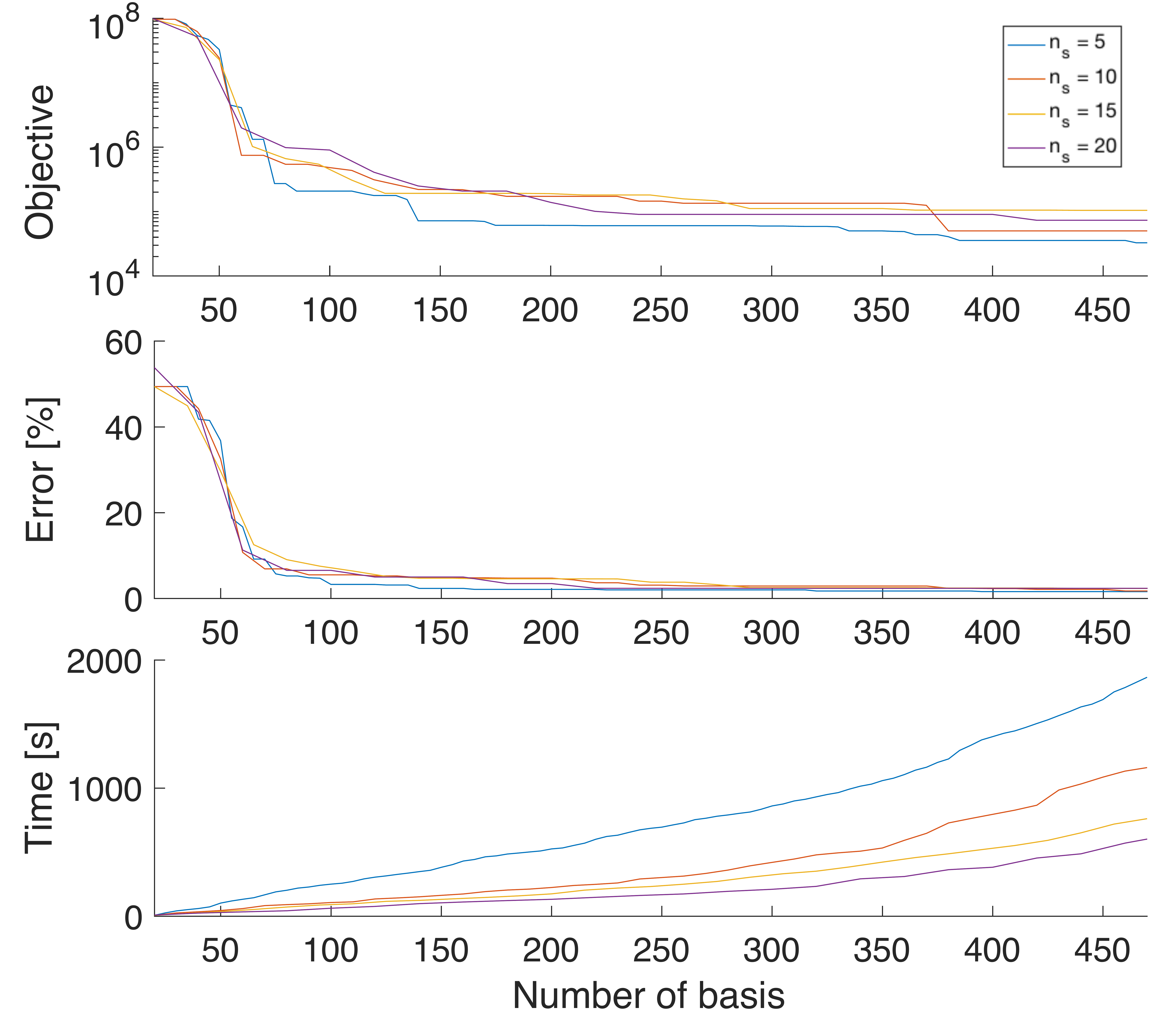}
  \caption{Effect of sliding amount to convergence with objective value (shown in log scale), target profile matching error and computation time.}
  \label{fig:slidingAmountEffect}
\end{figure}

Figure~\ref{fig:slidingAmountEffect} 
demonstrates convergence of our optimization with respect to number of explored 
basis functions for different sliding amount values, $n_s$. We provide objective value in log scale, the average percentage error in target profile matching and computation time. In terms of 
selecting the sliding amount, we found that lowering the sliding amount and 
re-optimizing for more basis functions might give better objective reduction 
capabilities for lower number of explored basis functions. However, as the 
number of explored basis increases, all $n_s$ values converge to similar 
objective values. Additionally, small sliding amounts take more computational 
time to cover the same amount of basis functions since more basis functions are 
re-optimized due to the overlap. In summary, we found that using large sliding 
amounts~($n_s ~ n_{opt}$) with small overlap works well resulting in similar 
objective minimization performance compared to small sliding amounts with minor 
increase in computational time. In our examples, we observed that $n_s \approx 
0.75 * n_{opt}$ gives a good trade off between objective minimization and 
computational performance.

\begin{figure}[]
  \centering  
  \includegraphics[trim = 0in 0in 0in 0in, clip, width = 0.9\columnwidth]{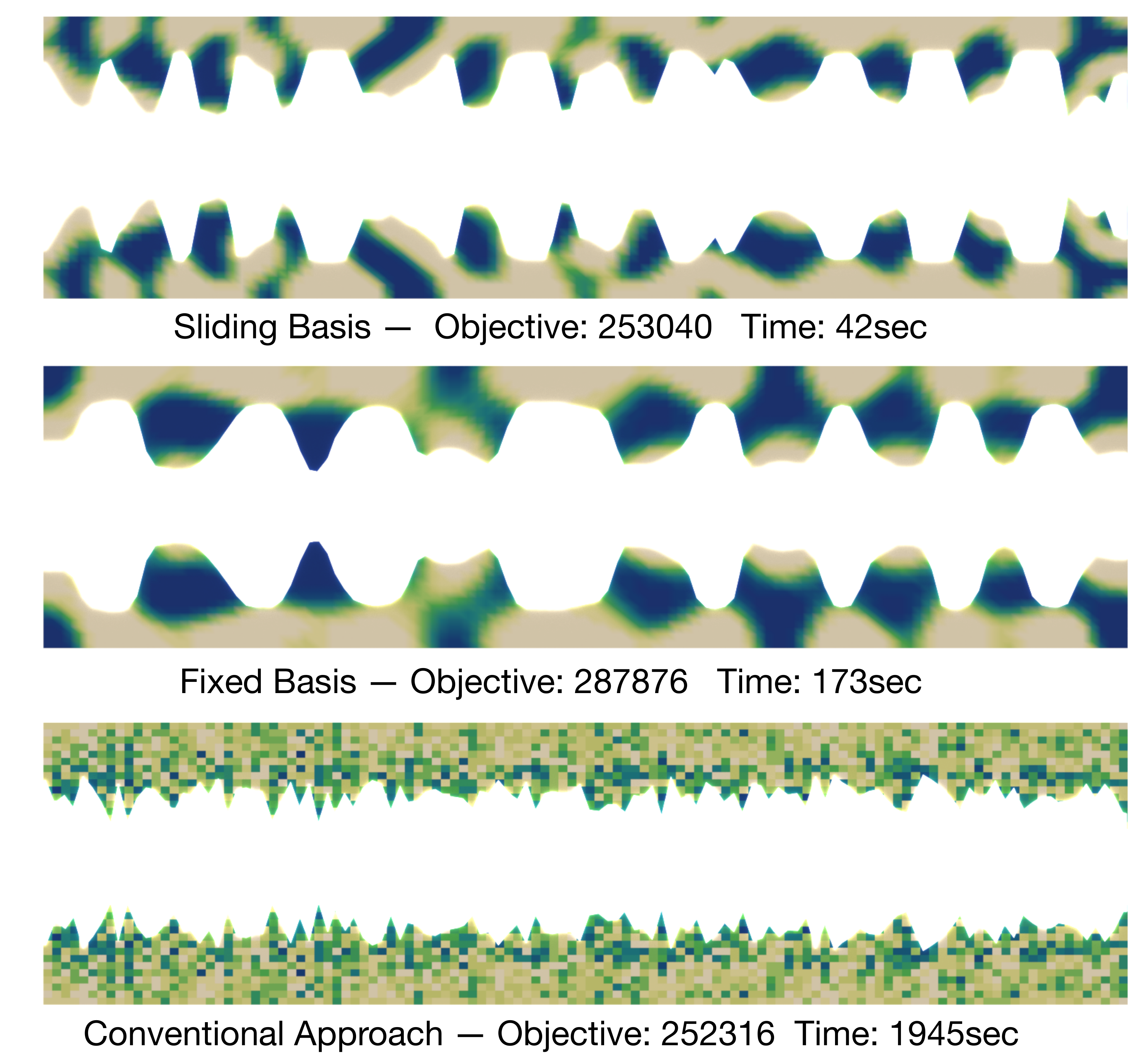}
  \caption{Comparison of objective value and computation time for sliding basis optimization, fixed basis optimization and conventional optimization approach.}
  \label{fig:rocketCompare}
\end{figure}

We compare our Laplacian basis formulation to classical topology optimization like conventional approach where value of element is optimized independently in Figure~\ref{fig:rocketCompare}. It can be observed that the results of the Laplacian basis are smoother since low number of basis functions are used and they correspond to lower frequency basis functions. We can see that there is less than 0.3\% objective value difference between the conventional approach and the sliding basis optimized result while the difference between the optimization time is significant. For this example, we run the solid rocket fuel design problem for the two-step thrust profile. Note that we have reduced grid resolution to 1800 to be able to run the conventional approach in practical times. As we increase the resolution and thus number of optimization variables, computational time gain increases.

\paragraph{Limitations} 

Since our formulation is a reduced order approach, the solution space is 
limited by the basis functions used during the optimization. Thus, there may be 
a better optimum that lies outside this space that cannot be reached by using a 
small number of basis functions ($k\sim100$) compared to optimizing for all 
elements of the mesh domain ($n_e\sim100k$). However, our examples 
demonstrate that the small number of basis functions are capable of satisfying 
the design requirements and enable solution of otherwise intractable 
computationally demanding problems.

We apply our sliding basis optimization to general nonlinear optimization 
problems. Unfortunately, finding the global optimum of such problems is still 
an open problem and our approach does not guarantee that the optimized 
solutions are globally optimum. Similar to traditional topology optimization 
approaches, different initial conditions may give different locally optimum 
results. Depending on the problem, this may be an important challenge. However, 
for compliance minimization, we do not observe significantly different results 
for different initial conditions in terms of the minimized compliance value. 
Moreover, our re-initialization at each step of the sliding optimization and 
re-optimization of overlapping basis functions help alleviate these local 
minima issues.

\section{Conclusion and Future Work}

Given a design domain and a set of goals in the form of objective and 
constraints, we present a method to design multi-material distributions. To 
solve the material distribution design problem with a reduced order approach, 
we use Laplacian eigenfunctions as a basis and project the design space into 
lower dimensions 
with fixed number of basis functions. Further, we extend this fixed basis 
optimization approach to a sliding basis method where the key idea is to 
exploit 
the spectral properties of the Laplacian basis for efficient exploration in the 
reduced space. Our sliding basis approach provides a flexible and powerful 
mechanism enabling computationally demanding design optimization problems 
involving black-box analysis. In this work, we show its efficiency on two 
applications as graded solid rocket fuel design and multi-material topology 
optimization. When tested with black-box analysis, our sliding basis approach 
can speed up the optimization process up to 8 times over the fixed basis method, 
often leading to better objective 
minimization due to local perturbations at each sliding step. We believe the 
ability to work with black-box analysis is very important for facilitating 
innovation and promoting engineering collaboration.

In this work, we focus on design material distribution fields. However, our 
approach generalizes to any spatial field design such as displacement fields. 
In future, other basis functions exhibiting similar spectral properties can be 
explored and incorporated into our sliding basis optimization method. Musialski 
\etal~\cite{Musialski:2015reduced} presents a reduced order method for surface 
design. Complementing our approach with a such parametrization could allow 
efficient design of surface geometries increasing surface detail in a sliding 
manner similar to increasing the detail in material distribution.

 \section{Acknowledgments}
The authors would like to thank NASA Jacobs Space Exploration Group for providing the solid rocket fuel design problem with the target thrust profile. This research was developed with funding from the Defense Advanced Research Projects Agency (DARPA). The views, opinions and/or findings expressed are those of the authors and should not be interpreted as representing the official views or policies of the Department of Defense or U.S. Government. 3D models:  dragon by XYZ RGB Inc and GE bracket by WilsonWong on GrabCAD.

\section*{References}
\bibliography{mybibfile}

\begin{thebibliography}{10}
\expandafter\ifx\csname url\endcsname\relax
  \def\url#1{\texttt{#1}}\fi
\expandafter\ifx\csname urlprefix\endcsname\relax\def\urlprefix{URL }\fi
\expandafter\ifx\csname href\endcsname\relax
  \def\href#1#2{#2} \def\path#1{#1}\fi

\bibitem{StratasysObjet}
{Stratasys} {Connex} {Objet},
  \url{https://www.stratasys.com/3d-printers/objet-350-500-connex3}, accessed:
  2019-08-07.

\bibitem{Insstek}
{Insstek} {DMT}, \url{http://www.insstek.com/content/multi_materials},
  accessed: 2019-08-07.

\bibitem{LiLiquin:2017}
L.~Li, J.~Wang, P.~Lin, H.~Liu, Microstructure and mechanical properties of
  functionally graded ticp/ti6al4v composite fabricated by laser melting
  deposition, Ceramics International 43~(18) (2017) 16638 -- 16651.
\newblock \href
  {http://dx.doi.org/https://doi.org/10.1016/j.ceramint.2017.09.054}
  {\path{doi:https://doi.org/10.1016/j.ceramint.2017.09.054}}.

\bibitem{Bandyopad:2018}
A.~Bandyopadhyay, B.~Heer, Additive manufacturing of multi-material structures,
  Materials Science and Engineering: R: Reports 129 (2018) 1 -- 16.
\newblock \href {http://dx.doi.org/https://doi.org/10.1016/j.mser.2018.04.001}
  {\path{doi:https://doi.org/10.1016/j.mser.2018.04.001}}.

\bibitem{Bendsoe:2004}
M.~P. Bends{\o}e, O.~Sigmund, Topology Optimization: Theory, Methods and
  Applications, Springer, 2004.

\bibitem{Davis:1991}
L.~Davis, Handbook of genetic algorithms.

\bibitem{Kirkpatrick:1983}
S.~Kirkpatrick, C.~D. Gelatt, M.~P. Vecchi, Optimization by simulated
  annealing, science 220~(4598) (1983) 671--680.

\bibitem{sorkine2005laplacian}
O.~Sorkine, Laplacian mesh processing, in: Eurographics (STARs), 2005, pp.
  53--70.

\bibitem{Zuo:2017multi}
W.~Zuo, K.~Saitou, Multi-material topology optimization using ordered simp
  interpolation, Structural and Multidisciplinary Optimization 55~(2) (2017)
  477--491.

\bibitem{Blasques:2014composite}
J.~P. Blasques, Multi-material topology optimization of laminated composite
  beams with eigenfrequency constraints, Composite Structures 111 (2014)
  45--55.

\bibitem{Wang:2004levelset}
M.~Y. Wang, X.~Wang, “color” level sets: a multi-phase method for
  structural topology optimization with multiple materials, Computer Methods in
  Applied Mechanics and Engineering 193~(6-8) (2004) 469--496.

\bibitem{mirzendehdel2015pareto}
A.~M. Mirzendehdel, K.~Suresh, A pareto-optimal approach to multimaterial
  topology optimization, Journal of Mechanical Design 137~(10) (2015) 101701.

\bibitem{Conlan:2019stress}
C.~Conlan-Smith, K.~A. James, A stress-based topology optimization method for
  heterogeneous structures, Structural and Multidisciplinary Optimization
  60~(1) (2019) 167--183.

\bibitem{Vogiatzis:2017multi}
P.~Vogiatzis, S.~Chen, X.~Wang, T.~Li, L.~Wang, Topology optimization of
  multi-material negative poisson’s ratio metamaterials using a reconciled
  level set method, Computer-Aided Design 83 (2017) 15--32.

\bibitem{Vaissier:2019graded}
B.~Vaissier, J.-P. Pernot, L.~Chougrani, P.~V{\'e}ron, Parametric design of
  graded truss lattice structures for enhanced thermal dissipation,
  Computer-Aided Design 115 (2019) 1--12.

\bibitem{Li:2018graded}
D.~Li, W.~Liao, N.~Dai, G.~Dong, Y.~Tang, Y.~M. Xie, Optimal design and
  modeling of gyroid-based functionally graded cellular structures for additive
  manufacturing, Computer-Aided Design 104 (2018) 87--99.

\bibitem{Choi:2019accelerating}
Y.~Choi, G.~Oxberry, D.~White, T.~Kirchdoerfer, Accelerating topology
  optimization using reduced order models, Tech. rep., Lawrence Livermore
  National Lab.(LLNL), Livermore, CA (United States) (2019).

\bibitem{Amsallem:2015}
D.~Amsallem, M.~J. Zahr, Y.~Choi, C.~Farhat, Design optimization using
  hyper-reduced-order models, Structural and Multidisciplinary Optimization 51
  (2015) 919--940.

\bibitem{Gogu:2015improving}
C.~Gogu, Improving the efficiency of large scale topology optimization through
  on-the-fly reduced order model construction, International Journal for
  Numerical Methods in Engineering 101~(4) (2015) 281--304.

\bibitem{Yoon:2010reduction}
G.~H. Yoon, Structural topology optimization for frequency response problem
  using model reduction schemes, Computer Methods in Applied Mechanics and
  Engineering 199~(25-28) (2010) 1744--1763.

\bibitem{Guest:2010reducing}
J.~K. Guest, L.~C. Smith~Genut, Reducing dimensionality in topology
  optimization using adaptive design variable fields, International journal for
  numerical methods in engineering 81~(8) (2010) 1019--1045.

\bibitem{poulsen2002topology}
T.~A. Poulsen, Topology optimization in wavelet space, International Journal
  for Numerical Methods in Engineering 53~(3) (2002) 567--582.

\bibitem{zhou2018highly}
P.~Zhou, J.~Du, Z.~L{\"u}, Highly efficient density-based topology optimization
  using dct-based digital image compression, Structural and Multidisciplinary
  Optimization 57~(1) (2018) 463--467.

\bibitem{white2018toplogical}
D.~A. White, M.~L. Stowell, D.~A. Tortorelli, Toplogical optimization of
  structures using fourier representations, Structural and Multidisciplinary
  Optimization 58~(3) (2018) 1205--1220.

\bibitem{Ulu:2018coupling}
N.~G. Ulu, S.~Coros, L.~B. Kara, Designing coupling behaviors using compliant
  shape optimization, Computer-Aided Design 101 (2018) 57--71.

\bibitem{Ulu:2017lightweight}
E.~Ulu, J.~Mccann, L.~B. Kara, Lightweight structure design under force
  location uncertainty, ACM Transactions on Graphics (TOG) 36~(4) (2017) 158.

\bibitem{levy2006laplace}
B.~Levy, Laplace-beltrami eigenfunctions towards an algorithm that"
  understands" geometry, in: IEEE International Conference on Shape Modeling
  and Applications 2006 (SMI'06), IEEE, 2006, pp. 13--13.

\bibitem{Levy:2002least}
B.~L{\'e}vy, S.~Petitjean, N.~Ray, J.~Maillot, Least squares conformal maps for
  automatic texture atlas generation, in: ACM transactions on graphics (TOG),
  Vol.~21, ACM, 2002, pp. 362--371.

\bibitem{Vallet:2008spectral}
B.~Vallet, B.~L{\'e}vy, Spectral geometry processing with manifold harmonics,
  in: Computer Graphics Forum, Vol.~27, Wiley Online Library, 2008, pp.
  251--260.

\bibitem{Liu:2004segmentation}
R.~Liu, H.~Zhang, Segmentation of 3d meshes through spectral clustering, in:
  12th Pacific Conference on Computer Graphics and Applications, 2004. PG 2004.
  Proceedings., IEEE, 2004, pp. 298--305.

\bibitem{sorkine:2004laplacian}
O.~Sorkine, D.~Cohen-Or, Y.~Lipman, M.~Alexa, C.~R{\"o}ssl, H.-P. Seidel,
  Laplacian surface editing, in: Proceedings of the 2004 Eurographics/ACM
  SIGGRAPH symposium on Geometry processing, ACM, 2004, pp. 175--184.

\bibitem{Xu:2015interactive}
H.~Xu, Y.~Li, Y.~Chen, J.~Barbi{\v{c}}, Interactive material design using model
  reduction, ACM Transactions on Graphics (TOG) 34~(2) (2015) 18.

\bibitem{Zhang:2004discrete}
H.~Zhang, Discrete combinatorial laplacian operators for digital geometry
  processing, in: Proceedings of SIAM Conference on Geometric Design and
  Computing. Nashboro Press, 2004, pp. 575--592.

\bibitem{Levy:2010course}
B.~L{\'e}vy, H.~R. Zhang, Spectral mesh processing, in: ACM SIGGRAPH 2010
  Courses, SIGGRAPH '10, ACM, New York, NY, USA, 2010, pp. 8:1--8:312.
\newblock \href {http://dx.doi.org/10.1145/1837101.1837109}
  {\path{doi:10.1145/1837101.1837109}}.

\bibitem{Song:2019multiscale}
R.~Song, L.~Wang, Multiscale representation of 3d surfaces via stochastic mesh
  laplacian, Computer-Aided Design 115 (2019) 98--110.

\bibitem{taubin1995signal}
G.~Taubin, A signal processing approach to fair surface design, in: Proceedings
  of the 22nd annual conference on Computer graphics and interactive
  techniques, ACM, 1995, pp. 351--358.

\bibitem{horak2013spectra}
D.~Horak, J.~Jost, Spectra of combinatorial laplace operators on simplicial
  complexes, Advances in Mathematics 244 (2013) 303--336.

\bibitem{goldberg2002combinatorial}
T.~E. Goldberg, Combinatorial laplacians of simplicial complexes, Senior
  Thesis, Bard College.

\bibitem{duval2002shifted}
A.~Duval, V.~Reiner, Shifted simplicial complexes are laplacian integral,
  Transactions of the American Mathematical Society 354~(11) (2002) 4313--4344.

\bibitem{spectra}
Y.~Qiu, {SpectrA}: {Sparse} {Eigenvalue} {Computation} {Toolkit} as a
  {Redesigned} {ARPACK}, \url{https://spectralib.org/} (2015--2019).

\bibitem{Nocedal:2006}
J.~Nocedal, S.~J. Wright, Numerical Optimization, Second Edition, Springer,
  2006.

\bibitem{Andreassen:2011top88}
E.~Andreassen, A.~Clausen, M.~Schevenels, B.~S. Lazarov, O.~Sigmund, Efficient
  topology optimization in matlab using 88 lines of code, Structural and
  Multidisciplinary Optimization 43~(1) (2011) 1--16.

\bibitem{Musialski:2015reduced}
P.~Musialski, T.~Auzinger, M.~Birsak, M.~Wimmer, L.~Kobbelt, Reduced-order
  shape optimization using offset surfaces., ACM Trans. Graph. 34~(4) (2015)
  102--1.

\end{thebibliography}

%
%
%
%

\end{document}